\documentclass[trackchanges]{aastex631}
\usepackage{bbm}
\usepackage{multirow}
\usepackage{amsmath}
\graphicspath{{./}{figures/}}

\begin{document}

\title{Identifying Light-curve Signals with a Deep-learning-based Object Detection Algorithm. II. A General Light-curve Classification Framework}

\correspondingauthor{Kaiming Cui}
\email{cuikaiming6@gmail.com}

\author[0000-0003-1535-5587]{Kaiming Cui}
\affiliation{Tsung-Dao Lee Institute, Shanghai Jiao Tong University, Shanghai, 201210, People's Republic of China}
\affiliation{Department of Physics, University of Warwick, Gibbet Hill Road, Coventry CV4 7AL, UK}
\affiliation{Centre for Exoplanets and Habitability, University of Warwick, Gibbet Hill Road, Coventry, CV4 7AL, UK}

\author[0000-0002-5080-4117]{D. J. Armstrong}
\affiliation{Department of Physics, University of Warwick, Gibbet Hill Road, Coventry CV4 7AL, UK}
\affiliation{Centre for Exoplanets and Habitability, University of Warwick, Gibbet Hill Road, Coventry, CV4 7AL, UK}

\author[0000-0001-6039-0555]{Fabo Feng}
\affiliation{Tsung-Dao Lee Institute \& School of Physics and Astronomy, Shanghai Jiao Tong University, Shanghai 201210, People's Republic of China}



\begin{abstract}

    Vast amounts of astronomical photometric data are generated from various projects, requiring significant effort to identify variable stars and other object classes. In light of this, a general, widely applicable classification framework would simplify the {process} of designing {specific} classifiers for {various astronomical objects}.
    We present a novel deep learning framework for classifying light curves using a weakly supervised object detection model. Our framework identifies the optimal windows for both light curves and power spectra automatically, and zooms in on their corresponding data. This allows for automatic feature extraction from both time and frequency domains, enabling our model to handle data across different scales and sampling intervals. We train our model on data sets obtained from Kepler, TESS, and Zwicky Transient Facility multiband observations of variable stars and transients. We achieve an accuracy of 87\% for combined variables and transient events, which is comparable to the performance of previous feature-based models. Our trained model can be utilized directly for other missions, such as the All-sky Automated Survey for Supernovae, without requiring any retraining or fine-tuning. To address known issues with miscalibrated predictive probabilities, we apply conformal prediction to generate robust predictive sets that guarantee true-label coverage with a given probability. Additionally, we incorporate various anomaly detection algorithms to empower our model with the ability to identify out-of-distribution objects.
    Our framework is implemented in the {\tt Deep-LC} toolkit, which is an open-source Python package hosted on Github (\url{https://github.com/ckm3/Deep-LC}) and PyPI.

\end{abstract}



\section{Introduction} \label{sec:intro}

A light curve (LC) is a time-dependent record of the intensity of light observed from astronomical objects. It contains valuable information about the object's properties and behavior. From an astrophysical perspective, the classification of LCs in various research fields primarily includes variable stars and transient sources.

Regarding variable star research, the classification can be quite diverse, as different types of stars exhibit unique characteristics due to different physical processes. This offers crucial perspectives into the structure and evolution of stars.
{Based on the General Catalog of Variable Stars \citep{samusGeneralCatalogueVariable2017} and data from the Gaia mission, the works of \citet{mowlaviGaiaDataRelease2018} and \citet{gaiacollaborationGaiaDataRelease2019} provide a comprehensive taxonomy tree of known variable stars.}
They categorize stellar variations, based on physical processes, into seven types: rotation, eclipse, microlensing, eruptive stars, cataclysmic sources, pulsation, and secular evolution. The more specific classification levels we usually use (such as eclipsing binary (EB), RS CVn, $\delta$ Sct, solar-like, etc.) all belong to subclasses of the above categories.

Like stellar sources, transients can also be regarded as a type of variable star with shorter time scales. According to the above classification, they should be categorized as eruptive stars. However, due to the different physical processes behind them, and the accompanying high-energy emission across multiple bands and messengers (such as X-ray, radio, neutrino, and gravitational wave), transients and variable stars are usually treated separately as different research areas.
Some typical transients are supernovae (SNe), cataclysmic variables (CVs), gamma-ray bursts, tidal disruption events, and fast radio bursts. In some studies, active galactic nuclei (AGN) are also classified alongside transients as extragalactic sources \citep{kesslerModelsSimulationsPhotometric2019}. Due to the nonrecurring nature of these events, rapid classification of transients is crucial. Immediate detection and multiwavelength follow-up are essential for a thorough understanding of the involved physical processes.

There are several citizen science projects, such as Citizen All-sky Automated Survey for Supernovae (ASAS-SN)\footnote{\url{https://www.zooniverse.org/projects/tharinduj/citizen-asas-sn}}, SuperWASP Variable Stars,\footnote{\url{https://www.zooniverse.org/projects/ajnorton/superwasp-variable-stars}} and Gaia Vari,\footnote{\url{https://www.zooniverse.org/projects/gaia-zooniverse/gaia-vari}} that aim to train the public in visually classifying LCs. These projects provide simple LCs and folded LCs with some auxiliary parameters, hoping to utilize human visual perception. However, due to the accumulation of a vast amount of photometric data, machine learning has become almost the only option for rapid and automatic LC classification. Recent computer vision algorithms can outperform human eyes in most cases. In terms of variable stars, recent representative space missions such as Kepler \citep{boruckiKeplerPlanetDetectionMission2010}, K2 \citep{howellK2MissionCharacterization2014}, and TESS \citep{rickerTransitingExoplanetSurvey2014} provide us with unprecedented high-precision and high-duty-cycle observational data. With this data quality, variable star classification can be performed well based on their distinct features with various classical machine learning techniques \citep[e.g.,][]{hinnersMachineLearningTechniques2018,armstrongK2VariableCatalogue2015,armstrongK2VariableCatalogue2016,audenaertTESSDataAsteroseismology2021,barbaraClassifyingKeplerLight2022}. Some vision-based convolution neural network (CNN) algorithms are utilized with the help of the frequency-domain information, mainly focusing on asteroseismology \citep[e.g.,][]{honDeepLearningClassification2018,honDetectingSolarlikeOscillations2018}. Ground-based observations also supply more general scenarios for exploring new algorithms. For example, {\citet{szklenarImagebasedClassificationVariable2020}} classify LCs with their folded LC images on Optical Gravitational Lensing Experiment data. \citet{naulRecurrentNeuralNetwork2018} employ a recurrent neural network (RNN) to alleviate the issue of nonuniformly sampled data to some extent. \citet{jamalNeuralArchitecturesAstronomical2020} test various time-series neural networks, including long short-term memory (LSTM), gated recurrent units (GRUs), temporal CNNs and dilated temporal CNNs.

Classifying ground-based multiband transient LCs is a challenging task. There are many algorithms that have been applied, including some feature-based classical algorithms \citep[e.g.,][]{villarSupernovaPhotometricClassification2019,
forsterAutomaticLearningRapid2021,sanchez-saezAlertClassificationALeRCE2021,sanchez-saezPersistentOccasionalSearching2023} which have achieved a high accuracy on both variables and transients from the Zwicky Transient Facility \citep[ZTF;][]{bellmZwickyTransientFacility2014,bellmZwickyTransientFacility2018}. Time-series algorithms based on deep learning techniques like CNNs, RNNs, GRUs
and LSTM are widely adopted \citep[e.g.,][]{charnockDeepRecurrentNeural2017,pasquetPELICANDeePArchitecturE2019,muthukrishnaRAPIDEarlyClassification2019,mollerSuperNNovaOpensourceFramework2020,beckerScalableEndtoendRecurrent2020}. Attention-based transformer algorithms have been actively explored in recent years \citep[e.g.,][]{pimentelDeepAttentionbasedSupernovae2022,allamjr.PayingAttentionAstronomical2023,donoso-olivaASTROMERTransformerbasedEmbedding2023,panAstroconformerProspectsAnalyzing2023}. There are also some methods that use a generative adversarial network \citep{garcia-jaraImprovingAstronomicalTimeseries2022}, as well as direct identification of photometric image sequences using a recurrent CNN model \citep{carrasco-davisDeepLearningImage2019}.

As time-domain surveys continue to evolve, higher cadences and larger fields of view are reducing the gap between observations of variable stars and transients. Classification algorithms also need to improve along this direction. In this work, we develop a general LC classification framework based on a weakly supervised object detection model. {A weakly supervised object detection involves training object detectors using only image-level labels for supervision. This approach differs from fully supervised methods, which use bounding boxes, and unsupervised learning, which operates without any labels \citep{zhangWeaklySupervisedObject2021}. Our classification framework} can extract time-domain and frequency-domain features as an imitation of human behavior. Section \ref{sec: dataset} describes our data set selection and preparation, Section \ref{sec: framwork} introduces our model architecture. Section \ref{sec: training&evaluation} shows the training process, hyperparameter choices, evaluation (Section \ref{sub: evaluation}), and uncertainty analysis (Section \ref{sub: uncertainty}). In Section \ref{sec: peculiar detection} we explore the anomaly detection with our model's output features. {Some limitations} of our model are discussed in Section \ref{sec: discussion}. {Finally, we summarize our work in Section \ref{sec: conclusion}, presenting our conclusions and discussing potential future applications.}

\section{Data Set Preparation} \label{sec: dataset}

A general classification framework should be able to handle different sources of data. First, space observations are typically more uniform and have fewer gaps, while ground-based observations are subject to inevitable daily windows but can offer longer baselines. Second, observations can be {broadband or multiband. Third, different missions have different target selections, noise properties, and systematics.} To cover both space- and ground-based observations, we choose Kepler and TESS as representatives of space telescope data, and ZTF as a representative of ground-based observations. To have a reliable evaluation of our model, we adopt the same training data from previous state-of-the-art works. Our Kepler training sample is from the recent TESS Data for Asteroseismology classification algorithm \citep[T'DA;][]{audenaertTESSDataAsteroseismology2021},\footnote{\url{https://github.com/tasoc/starclass}} and we adopt the Kepler ID catalog but select data from all available quarters. We obtain our TESS training samples from the visual inspection conducted by \cite{balonaIdentificationClassificationTESS2022}. {We utilize the initial four years (Sector 1--55) of 2\,minute data from the TESS Science Processing Operations Center \citep{jenkinsTESSScienceProcessing2016}}. Presearch data conditioned simple aperture photometry \citep[PDCSAP;][]{smithKeplerPresearchData2012} LCs are used for both Kepler and TESS samples. {To enhance the generality of our model, we minimize the preprocessing of the LCs. For both Kepler and TESS LCs, we concatenate all accessible quarters or sectors. This concatenation entails a straightforward stitching process with normalized flux for each quarter or sector. The normalization is $(f - \text{med}(f)) / |\text{med}(f)|$, $f$ is the flux and $\text{med}$ is the median function. We do not discard outliers in the LCs because our classifier is designed for both variables and transients, which may have different outlier thresholds. Moreover, disparate systematics and varying observational precision necessitate different outlier thresholds. Instead, we allow our model to autonomously learn to manage outliers.} Kepler and TESS data are processed with {\tt lightkurve} \citep{lightkurvecollaborationLightkurveKeplerTESS2018,geert_barentsen_2021_4603214} and our extension \citep[{\tt lightkurve-ext};][]{cuiLightkurveextExtensionLightkurve2024}, which is available on GitHub\footnote{\url{https://github.com/ckm3/lightkurve-ext}}. All the Kepler and TESS data used in this paper can be found in MAST \citep{stsciKeplerLCQ0Q172016,mastteamTESSLightCurves2021}. ZTF training data set is downloaded from the Automatic Learning for the Rapid Classification of Events (ALeRCE) broker \citet{sanchez-saezAlertClassificationALeRCE2021}\footnote{\url{https://zenodo.org/doi/10.5281/zenodo.4279622}}.

Our combined training data set is formed from those three sets, while combining their taxonomy. {The combined taxonomy is listed in Table \ref{dataset-table}. Our taxonomy of variable stars follows T'DA. The $\delta$ Sct and $\beta$ Cep are merged as they have {similar} pulsation mechanisms and pulsation period ranges, causing {similar} characteristics on LC. RR Lyrae and Cepheid stars are combined as they are both radial pulsators and have some period overlaps. For some common classes in T'DA and ALeRCE, we also make adjustments to ensure there is no conflict among their definitions.} First, {we exclude the aperiodic class defined by T'DA, which selects long-period variables (LPVs) from Kepler exhibiting periods longer than 30\,days. This decision is made in order to simulate the aperiodic signals detected within a single TESS sector. We do not combine those selected LPVs with the ALeRCE LPVs because they can bias the long-period rotators \citep{cuiLongRotationPeriod2019} without  knowledge of the stellar parameters (e.g., surface gravity $\log g$). A more comprehensive analysis of LPVs that includes stellar parameters may yield refined classifications; we intend to address this in future research endeavors.} Second, the Periodic-Other sample in ALeRCE data set is a catchall-type class, summarizing periodic variable stars without a clear label. However, this can cause confusion with some classes {(e.g., rotation and contact EBs)} in the T'DA data set, so it is excluded from our data set. Then, because the `Constant' type in T'DA is only designed for TESS, we develop a new class named ``Random,'' which purely generates white noise. This noise flux follows a normal distribution $\mathcal{N}(0, \sigma)$, and the $\sigma$ follows log-uniform distribution $\mathcal{LU}(-6, 1)$ with base 10. The noise time cadences are sampled uniformly between 1\,minute and 2\,hours, and the time span uniformly from 10 to 1400\,days, generally following the Kepler and TESS observation cadences and lengths.

Finally, our combined data set contains 17 different classes, with a total of 24,552 objects and they are summarized in Table \ref{dataset-table}. Since we do not want to place emphasis on either variable (Kepler/TESS) or transient (ZTF) data, we undersample certain classes to maintain a nearly equal size of the ZTF and Kepler/TESS data sets while avoiding a reduction in less frequent categories. Usually variable stars are easier to classify, more variables can bias our performance, and a smaller total sample size can reduce our training time significantly.
We do not employ any imbalanced learning technique to oversample less represented classes, as it may not always be beneficial and could potentially harm our model's calibration \citep{elorSMOTENotSMOTE2022}.
{We also select a data set of variable stars that solely contain LCs from Kepler and TESS. It includes first seven classes of Table \ref{dataset-table}. This data set is created to facilitate a comparison with the performance of T'DA.}
The training and test data sets are divided at a ratio of 80\% and 20\%, respectively. The validation data set is made by fivefold cross-validation from the training data. Note that our sample is mainly designed for maintaining consistency with previous work; more complete samples from different missions and simulations can be collected in the future (see Section \ref{sec: conclusion}).

\begin{deluxetable}{cC}
    \tablecaption{Summary of the data set \label{dataset-table}}
    \tablehead{\colhead{Class} & \colhead{Number ($N_\mathrm{Kepler} + N_\mathrm{TESS}$ + $N_\mathrm{ZTF}$)} }
    \startdata
    Rotation or contact EBs                                          & 4113 (2260 + 1853 + 0)  \\
    $\delta$ Sct or $\beta$ Cep Stars                          & 3057 (772 + 1553 + 732)   \\
    $\gamma$ Dor or slowly pulsating B stars                     & 1746 (630 + 1116 + 0)   \\
    Random\tablenotemark{a}                                                           & 1000     \\
    RRL or Cephedi stars                                       & 1855 (43 + 194 + 1618)    \\
    Solar-like pulsators                                            & 1811 (1800 + 11 + 0)  \\
    Eclipses or transits                              & 1974 (974 + 0 + 1000)  \\
    AGN                                                              & 1000 (0 + 0 + 1000)   \\
    Beamed-jet-dominated AGN (blazars)                                                           & 1267 (0 + 0 + 1267)  \\
    CVs/novae                                                        & 871  (0 + 0 + 871)   \\
    LPVs                                                              & 1400 (0 + 0 + 1400)\\
    Type 1 quasars (QSOs)                                              & 1000 (0 + 0 + 1000)   \\
    Superluminous SNe                                  & 24 (0 + 0 + 24)  \\
    SNe II                                                            & 328 (0 + 0 + 328)    \\
    SNe Ia                                                            & 1272 ( 0 + 0 + 1272)\\
    SNe Ibc                                                           & 94 ( 0 + 0 + 94)   \\
    Young stellar objects                                       & 1740 (0 + 0 + 1740)\\  
    {Total}                                       & {24,552 (6479 + 4727 + 12,346 + 1000\tablenotemark{b})}
    \\
    \enddata
    \tablenotetext{a}{{The Random class is synthetic white noise LCs with no variability.}}
    \tablenotetext{b}{{The synthetic Random class does not belong to any survey.}}
\end{deluxetable}

\begin{figure}
    \centering
    \includegraphics[width=\textwidth]{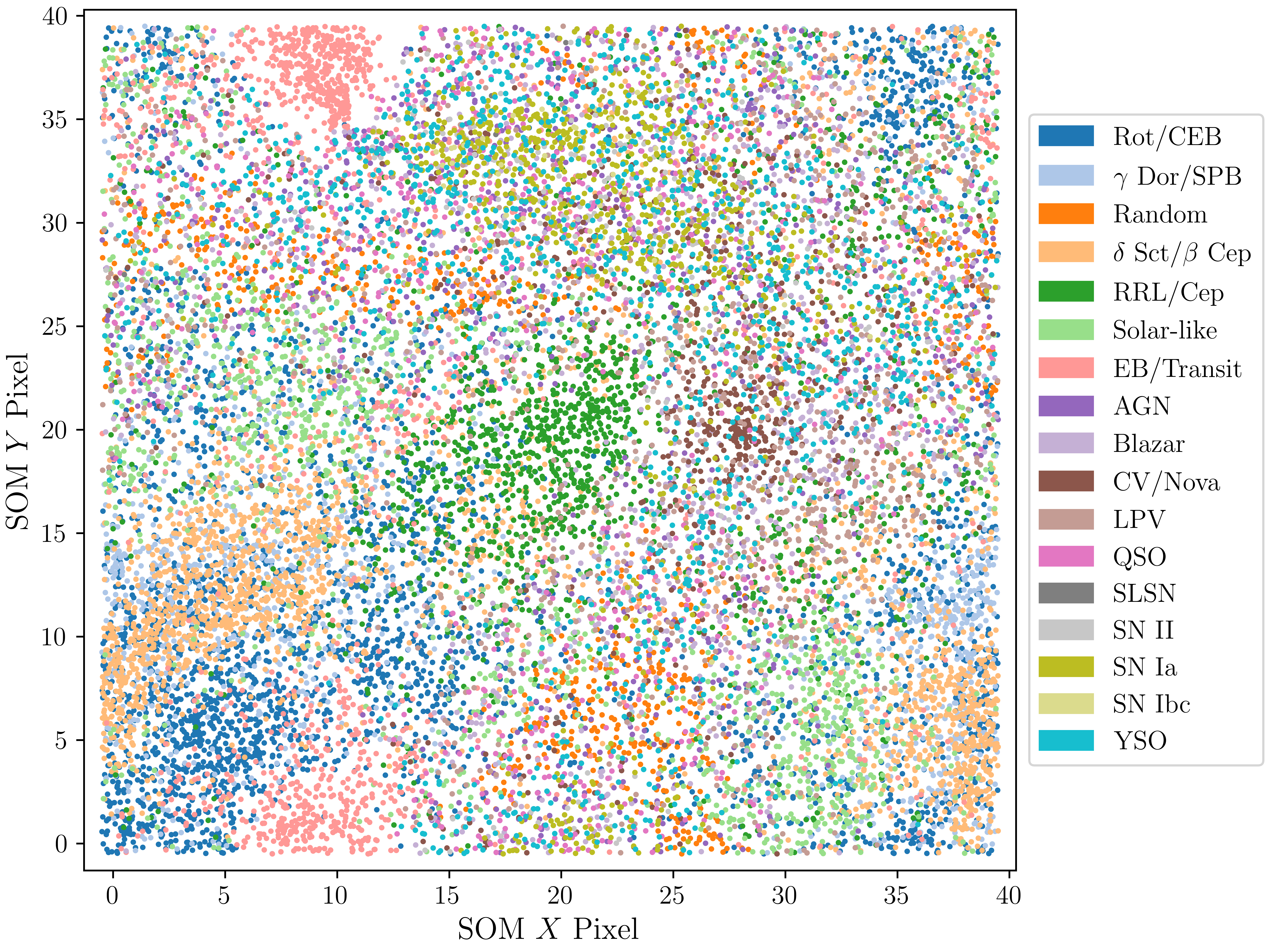}
    \caption{SOM 2D representation of the training data set. {Random jitter within each pixel has been added for clarity. Each colorful point represents a LC in our training data set; the different colors correspond to different classes as denoted in the legend.}}
    \label{fig:som}
\end{figure}

To explore the distribution of our data set and also investigate the difficulty level of classification, we train a self-organizing map \citep[SOM;][]{kohonenSelforganizingMap1990} model {with} our training data set. We {simply} use the folded LC based on the most significant frequency, following the methodology of \citet{armstrongK2VariableCatalogue2016}. {As an unsupervised learning algorithm, SOM projects different classes onto a 2D output plane according to their features, ensuring that similar classes are positioned in close proximity on the SOM grid. Therefore, the} typology of folded LC features {in} our {training data set} is shown in Figure \ref{fig:som} {using} trained SOM results. {Within a given category, LCs are expected to cluster within a similar SOM pixel region, whereas distinct categories should be well separated.} Some labels, such as EB and RRL, can be easily classified and set apart from others, but {significant overlap is observed among most classes}. This demonstrates that a simple phase-folded LC is insufficient for classifying all these diverse classes.

\section{Classification Framework} \label{sec: framwork}
An ideal scenario for time-series analysis is to have a network that utilizes a single algorithm capable of extracting all relevant information from both the input time-series data points and the input sample distribution. With such a network, it would be possible to complete all tasks related to time-series analysis in an end-to-end manner. However, even sophisticated transformer models have not managed to surpass the performance of straightforward linear models on some real-life data sets \citep{zengAreTransformersEffective2022}. Furthermore, enhancing or integrating traditional algorithms can potentially lead to superior outcomes \citep{sunFreDoFrequencyDomainbased2022}. Instead of relying on a single time-series model, our model integrates domain knowledge. This will allow the designing of distinctive feature extraction modules tailored to the physical properties of each category.

Inspired by previous work on visually fine-grained classification and validation of LCs \citep[e.g.,][]{balonaKeplerObservationsScuti2011,gaulmeSystematicSearchStellar2019,twickenKeplerDataValidation2018}, the starting point for classification is usually the morphology of LCs at different time scales, as well as frequency-domain information and some auxiliary physical parameters.
Thus, unlike most time-series models, a general model should utilize time-domain, frequency-domain and physical parameters all together. Additionally, it must possess the capability to independently manipulate or process data in order to adapt to varying data scenarios.
Most classification models that use LCs as input are usually designed for specific missions, with predetermined lengths of LC inputs and known observation cadences and colors. To enable our model to handle such diverse inputs, we apply various techniques to mitigate this issue.

First, in order to allow the model to handle variable-length and nonequal-interval sampled sequence data, we convert the LC to images, together with the original LC data, to serve as the input for the network.
{Our input LC to image conversion follows the similar process in \citet{cuiIdentifyLightcurveSignals2022} with some modifications. We draw 1D LC sequence data to a 2D scatter point image, and the input image shape is (1, 256, 512), where the first dimension is the number of channels. The second and third dimensions are the height and width, respectively. The scatter points are squares with 5 pixel sizes. For ZTF LCs with {\it r} and {\it g} bands, to distinguish those two bands, the scatter points are drawn with {different} grey scales (that range from 0 to 1), 1 for the {\it r} band, which is pure black, 0.5 for the {\it g} band. Also, their sizes are 4 and 5 pixels. The LC to image conversion} has several advantages. Initially, models for 1D time-series such as 1D CNN require a fixed shape and other models like RNNs and LSTM have limited perception range, making them difficult to process inputs spanning multiple orders of magnitude (such as TESS's 2\,minute cadence 27.4\,day observations and daily ground-based long-term observations). Previous preprocessing techniques such as binning, padding, and interpolation can alter the flux information for different timescales or result in redundancy. However, for 2D images, the data size remains unchanged while the flux and time range information is not affected by resolution. Additionally, observation data from different bands can be naturally represented by different colors without any requirement for data size and time alignment. Moreover, image data also allows us to fully utilize well-established 2D CNN models that are widely used in computer vision applications. Some studies also apply this method for LC or sequential data analysis, which benefits from fixed-size inputs and leveraging existing 2D CNN algorithms. {\citep[e.g.,][]{mahabalDeeplearntClassificationLight2017,honDetectingSolarlikeOscillations2018,szklenarImagebasedClassificationVariable2020,claytorRecoveryTESSStellar2022,cuiIdentifyLightcurveSignals2022}}. Furthermore, converting time-series into images can be regarded as a type of data {embedding} that integrates human perception {and prior knowledge}. This approach is highly intuitive and facilitates the interpretation of model outcomes.
However, since images only provide compressed representations at certain scales, the data input accepted by models is severely affected by factors such as image size and resolution. Therefore, we have to integrate zooming capabilities into our model to allow for analysis at smaller scales, simulating human behavior in LC classification. {Many exoplanet vetting works apply preset zoomed-in regions (local views) of the LC to aid the classification \citep[e.g.,][]{shallueIdentifyingExoplanetsDeep2018,yuIdentifyingExoplanetsDeep2019,valizadeganExoMinerHighlyAccurate2022}. A fixed zoomed-in region is possible for a specific class, but a general classification model needs to have a more flexible zooming ability.}

Developing a neural network algorithm with the ability to zoom is {challenging} because it requires the network to have the capability of perception {(determining regions for zooming) and action (executing the zoom process)}. For perception, experiments show that different layers of the CNN model will activate corresponding features for the target, which is also the basis of object detection algorithms \citep{zeilerVisualizingUnderstandingConvolutional2014}. Based on activation regions from feature maps, reinforcement learning seems like a straightforward approach for learning how to perform zooming operations. This involves rewarding or penalizing the network's zooming behavior based on classification results in the zoomed region \citep[e.g.,][]{gaoDynamicZoominNetwork2018}. However, we find that reinforcement learning is difficult to converge in our classification task. The main reason could be challenges in establishing a relationship between the reward function and the classification scores. Additionally, due to the sparsity of the action space, the efficiency of the parameter updating is low, leading to decreased training efficiency.

Another option is weakly supervised object detection methods, which detect and localize instances with solely image-level classification labels \citep{shaoDeepLearningWeaklySupervised2021}. {With this kind of algorithm, the zooming process can be naturally applied to the localized regions.} In this work, we adopt the algorithm from a fine-grained classification and object detection model called the Navigator-Teacher-Scrutinizer Network \citep[NTS-Net;][]{yangLearningNavigateFinegrained2018}. This model employs a weakly supervised learning technique to achieve simultaneous {detection and classification}. The main process of this algorithm involves using a navigator network to propose regions of interest from the image, and {the teacher/scrutinizer network makes predictions based on features extracted from the proposed regions. Both teacher and scrutinizer networks share the same inputs, but while the teacher network classifies the raw image and each zoomed-in region, the scrutinizer combines these features to make a final prediction.} The navigator and the {teacher/scrutinizer} work together in an iterative manner, refining the regions of interest and improving the accuracy of the detection, and better detection returns higher classification scores. The navigator network applies anchor boxes and a region proposal network \citep[RPN;][]{renFasterRCNNRealTime2016} as an object detector, {utilizing a pretrained CNN model for feature extraction. The teacher/scrutinizer network then provides classification scores based on the features extracted from the proposed regions.}
The model's primary loss is composed of a few distinct losses: the cross-entropy loss, which serves as the fundamental loss for classification, is applied to the raw LC, zoomed-in LCs, and final combined data; a rank loss is applied to align the sorting of the output scores from the RPN with the sorting of the correct classification logits. By doing so, the network achieves a weakly supervised object detection process.

\begin{figure}
    \centering
    \includegraphics[width=\textwidth]{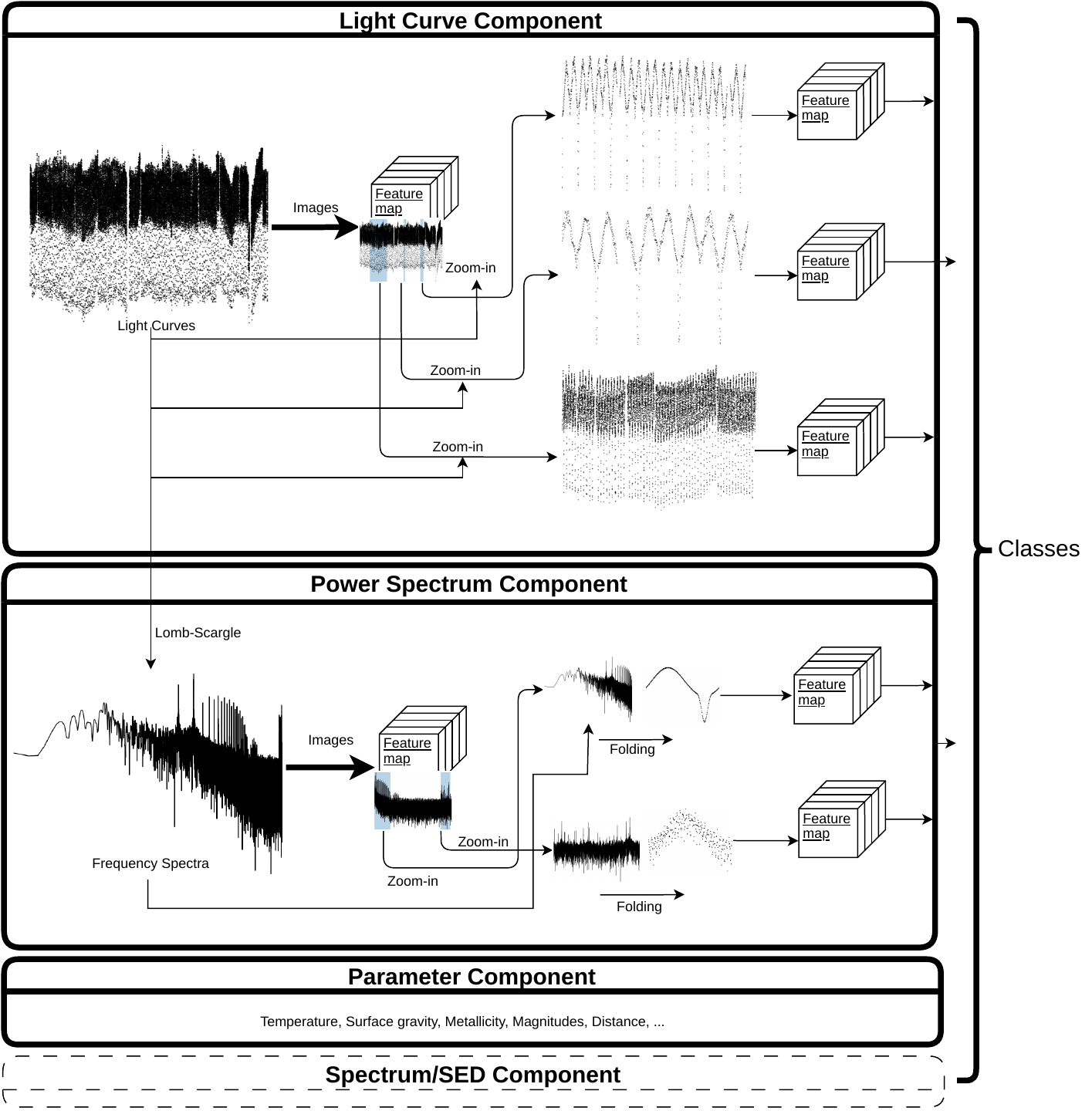}
    \caption{Network schematic diagram of our model. The diagram shows three main components, which are enclosed by thick black lines. At the top of these components is the LC component, which has two inputs: raw LC data and image data of plotted LCs. A CNN model extracts feature maps from the images and zooms in on the input LC based on these feature maps. Then, the zoomed-in LCs enter into the same CNN model to extract features again. Finally, all {the extracted} features are combined to make a prediction for the class label. The PS component receives the LC from the LC component and calculates a Lomb--Scargle periodogram as input for the network. Similar to the LC component, the frequency spectra and their images are feature-extracted and zoomed in to smaller regions. Then, with the most significant frequencies in the zoomed-in regions, {we can phase-fold the LCs}, which are then used by a CNN model to extract features. The predictive labels are generated by combining all of the features produced by these CNN models. The parameter component utilizes all manually selected physical or systematic parameters to make higher-level predictions. The optional spectrum or SED component demonstrates the potential for extensibility to additional astronomical products.}
    \label{fig:architecture}
\end{figure}

After the zooming capability is implemented in our model, it can effectively analyze LCs of different scales and determine appropriate scales for classification. Figure \ref{fig:architecture} illustrates the structure and process of our model, which includes an LC component responsible for feature extraction from LC images and zooming in on LC data. The final label prediction is generated by integrating both raw and zoomed-in LC data. We have also implemented a similar process for the power spectrum (PS), which is calculated with a Lomb--Scargle algorithm \citep{lombLeastSquaresFrequencyAnalysis1976, scargleStudiesAstronomicalTime1982}, allowing it to accommodate PSs with various frequency resolutions and ranges. {For the multiband LC, we choose the multiband Lomb--Scargle implemented in \citet{vanderplasPeriodogramsMultibandAstronomical2015}.} The minimum frequency is two times the Rayleigh
resolution ($1/\Delta T$, where $\Delta T$ is the time length), while the maximum frequency is the Nyquist frequency, calculated based on the minimum cadence. The frequency is oversampled by a factor of 5. Slightly different from the LC component, in the PS component, we use the zoomed-in region to identify significant frequencies and then fold the original LC according to these frequencies. We combine the folded LCs with the original spectrum for the final classification prediction. In fact, for the PS component, this is a more efficient approach because unlike that with an LC, the specific frequency (location information) holds great importance and does not have shift invariance. {The LC and PS} components share the same input LC and can work independently or together for comprehensive analysis. {The model can also output the intermediate classification results for each region that has been zoomed-in.} In addition to these two components, Figure \ref{fig:architecture} also displays the parameter component, which can accept predetermined astrophysical parameters to make finer classifications. We also show the design of the spectrum/spectral energy distribution (SED) component; by applying similar techniques, our model will have the potential to expand to more types of astrophysical observation data. Two additional intuitive animations that demonstrate zooming and classification for both the LC and PS components are accessible on our package's website.\footnote{\url{https://github.com/ckm3/Deep-LC}}

\begin{figure}
    \centering
    \includegraphics[width=0.9\textwidth]{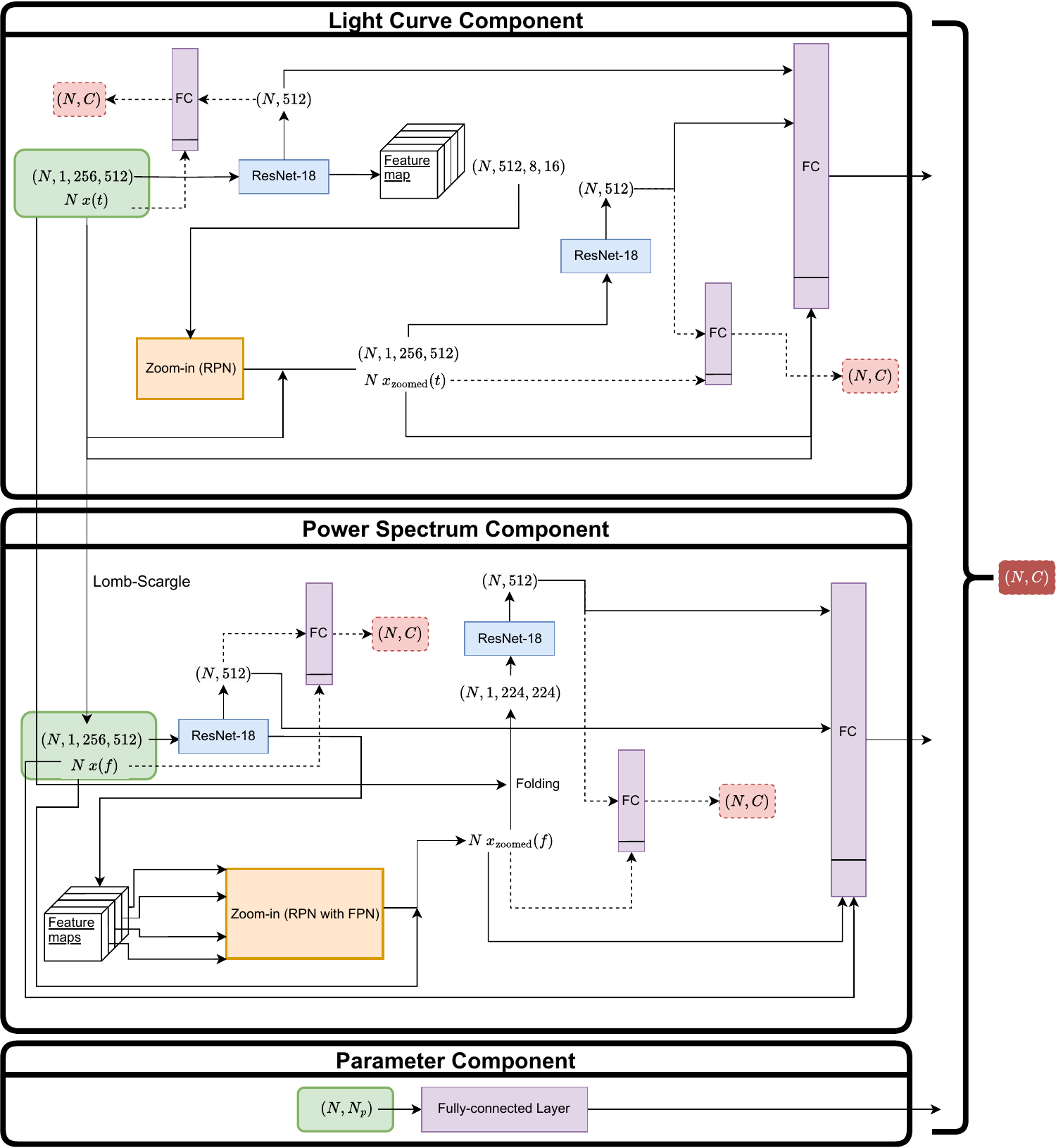}
    \caption{Details of the network structure's implementation, as depicted in Figure \ref{fig:architecture}. {Similar modules in our model are coded with the same colors. The LC component comprises the input for $N$ LCs ($x(t)$) and images with a shape of batch size, channel number, height, and width, as illustrated within the solid rounded {green} rectangle. ResNet-18 can generate feature maps both with and without spatial information. The RPN module is shown as an orange box, while the zoomed-in LCs are represented as $x_\mathrm{zoomed}(t)$. The FC layer is divided into two sections: the lower portion represents the input of three basic LC features, and the upper portion signifies the input feature map. The output of the FC layer has a shape of ($N$, $C$), where $C$ denotes the number of classes. By integrating all of the zoomed-in and initial features, the final FC layer produces class predictiosns. Dashed connectors and outputs are optimized during training and can be utilized as needed during inference. For the PS component, inputs include $N$ PSs ($x(f)$) and images computed directly from LCs; this process mirrors that of the LC component. Lastly, the parameter component processes the preset parameters with an FC layer. $N_p$ is the {number} of parameters of each object.}}
    \label{fig:architecture-net}
\end{figure}

During the implementation of our model, we extensively modify NTS-Net to suit our LC classification task.
The first and most direct adjustment is {on the navigator network. We zoom in on the 1D time-series to ensure no loss of information rather than} employ 2D image cropping.
Secondly, relying solely on ResNet classification for images is insufficient. To enhance the model's understanding of LCs with varying time scales, amplitudes, and sampling rates, necessary information such as the total time length, maximum variability, and minimum cadence is incorporated into fully connected (FC) layers {of the teacher/scrutinizer network. Those pieces of variability and time scale information} act like the coordinates of the LC, essential for the model to understand LCs with different time scales, amplitudes, and sampling rates. {LCs are normalized so that relative variability can be directly extracted.}
Third, in contrast to image resizing, our zooming only requires manipulation in the time dimension; therefore, {in the navigator network,} we replace the 2D anchor boxes with 1D anchor scales.
Moreover, {as in most object detection algorithms, we need to apply nonmaximum suppression \citep{neubeckEfficientNonMaximumSuppression2006} to eliminate duplicated regions from RPN, and only keep the most relevant regions. However,} even after applying nonmaximum suppression on the image level, it is still possible for an LC or PS to return duplicate regions {if the zoomed-in regions contain gaps or edges}. Thus we further detect and remove duplicate LCs or PSs from the candidate regions proposed by the RPN.
To achieve a balance between computational cost and classification accuracy, we utilize a relatively lightweight ResNet-18 \citep{heDeepResidualLearning2015} model as the backbone for feature extraction and image classification. {The detailed data flow of our model is shown in Figure \ref{fig:architecture-net}. LCs and PSs exhibit distinct data characteristics.} When using the RPN to provide candidate zooming regions, LCs and PSs require different strategies for region generation.

\begin{figure}
    \centering
    \includegraphics[width=0.8\textwidth]{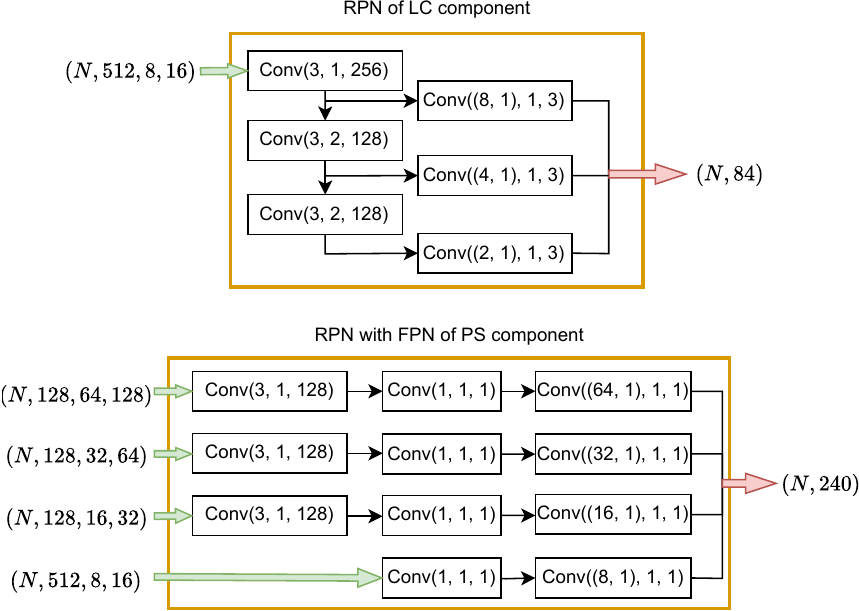}
    \caption{{Details of the RPN structure for the LC and PS components.  From left to right, we present the input to output flow. The shape of the input is defined by batch size, number of channels, height, and width. The numbers in the convolutional blocks represent filter sizes (shapes), strides, and the number of output channels.}}
    \label{fig:architecture-RPN}
\end{figure}

The LC sample contains data points and time lengths spanning several orders of magnitude, so there should be more scaling sizes of region candidates to cover different time scales{. However, variations in the region's position can be considered negligible for classification purposes. In contrast, successful classification of PS data depends on particular frequency values or some identifiable patterns. Only when the proposed zooming region encompasses these critical frequency bands effectively can the classfication be considered valid.}
Previous experiments observed that the shallow layers of a CNN model primarily respond to edges and textures, whereas deeper layers exhibit more class-specific semantic information \citep{zeilerVisualizingUnderstandingConvolutional2014}. Thus to optimize the performance of object detection in both LC and PS components, we propose different RPNs as illustrated in {Figure \ref{fig:architecture-RPN}. An RPN integrates several CNN blocks and is capable of providing scores at each pixel on the feature map. Specifically, the RPN for the LC component only processes the final feature layer of ResNet (feature map shown in Figure \ref{fig:architecture-net}), while the RPN for PS component utilizes the last four layers of ResNet (feature maps shown in Figure \ref{fig:architecture-net}) to construct a feature pyramid network \citep[FPN;][]{linFeaturePyramidNetworks2017} and return more region candidates. The FPN is a prevalent method that augments detection accuracy by merging features from various levels, thereby facilitating accurate frequency localization in PS data.
In our implementation, we have predefined anchor regions for both the LC and PS components. Within the LC component's RPN feature map, each pixel is associated with nine anchor regions, with sizes ranging from 32 to 384\,pixels. A larger number of anchors translates to a broader range of potential scaling options. Conversely, the PS component's RPN feature map contains four anchor regions per pixel, with scales varying from 10\% to 40\% of the original image size.}

In addition, we have a few limits for region selection. {In \citet{sanchez-saezAlertClassificationALeRCE2021}, the full LCs have at least six data points. Therefore, in our} LC component, when the selected LC has less than six data points, the selection is ignored. In the PS component, we calculate the false-alarm probability (FAP) levels on the whole PS {based on the method of \citet{baluevAssessingStatisticalSignificance2008}} and discard regions with FAP levels lower than 1\% in subsequent processes. For multiband observations, bootstrap simulation is required to calculate FAP and this process is computationally expensive. Therefore, it has been disabled in our ZTF data set but can be enabled in the inference with our package.
For observations with large gaps, we aim to limit the model's focus to regions with more data points. To achieve this, we introduce a new loss function that penalizes network outputs for blank areas only. This is expressed in Equation \ref{eq: loss0}:
\begin{equation}
    L_\mathrm{0} = \mathbbm{1}^\mathrm{<0}  \left( \frac{1}{N_\mathrm{\ne 0}} \sum_{i=0}^{N_\mathrm{\ne 0}} S_i \mathbbm{1}_i^\mathrm{\ne 0} - \frac{1}{N_\mathrm{=0}} \sum_{i=0}^{N_\mathrm{=0}} S_i \mathbbm{1}_i^\mathrm{=0} \right) \label{eq: loss0}
\end{equation}
$S_i$ is the score given by the RPN, $N_\mathrm{=0}$ and $N_\mathrm{\ne0}$ are the number of candidate zoomed-in regions with blank and nonblank regions, and $\mathbbm{1}_i^\mathrm{\ne0}$ and $\mathbbm{1}_i^\mathrm{=0}$ are the Heaviside step functions for nonzero and zero regions, respectively. The function $\mathbbm{1}_i^\mathrm{\ne0}$ equals 1 when the $i$th region is not blank, else 0; $\mathbbm{1}_i^\mathrm{=0}$ equals 1 when the $i$th region is is blank. The $\mathbbm{1}^\mathrm{<0}$ selects the loss less than 0, making sure the gradient is only back-propagated to the blank areas.

Therefore, the final loss function of our LC or PS component is in the form of Equation \ref{eq: loss_total}:
\begin{equation}
    \label{eq: loss_total}
    L = L_\mathrm{C} + L_\mathrm{S} + L_\mathrm{ResNet} + L_\mathrm{Rank} + L_\mathrm{0},
\end{equation}
where $L_\mathrm{C}$ is the loss function of the final combined layer, and $L_\mathrm{S}$ is the loss function from the scrutinizer part, which combines the LCs or PSs from the proposal regions. $L_\mathrm{ResNet}$ is the initial loss from the raw images. The above three loss functions all use cross-entropy losses to minimize  differences between the predictions and ground-truth labels. {They are listed as follows:}
\begin{equation}
    L_\mathrm{C} = -\sum_{i=1}^{N} \sum_{c=1}^{C} y_{C_{i,c}} \log(p_{C_{i,c}}), \label{eq: loss_C}
\end{equation}
{where $y_{C_{i,c}}$ is the ground-truth label, $p_{C_{i,c}}$ denotes the predicted probability for the combined layer output (this predicted probability is computed using the softmax function applied to the output logits, expressed as $p_{C_{i,c}} = \exp(x_{i,c})/{\sum_{c=1}^{C}\exp(x_{i,c})}$), $N$ is the batch size, and $C$ is the number of classes.}
\begin{equation}
    L_\mathrm{S} = -\sum_{i=1}^{N} \sum_{c=1}^{C} y_{S_{i,c}} \log(p_{S_{i,c}}), \label{eq: loss_S}
\end{equation}
{where $y_{S_{i,c}}$ is the ground-truth label, and $p_{S_{i,c}}$ is the predicted probability from the scrutinizer, which includes the output of the proposed regions. The predicted probability is also calculated using the softmax function on the output logits.}
\begin{equation}
    L_\mathrm{ResNet} = -\sum_{i=1}^{N} \sum_{c=1}^{C} y_{R_{i,c}} \log(p_{R_{i,c}}), \label{eq: loss_ResNet}
\end{equation}
{where $y_{R_{i,c}}$ is the ground-truth label, and $p_{R_{i,c}}$ is the predicted probability from the ResNet backbone, which is the loss of the input raw image, also calculated using the softmax function on the output logits.}
$L_\mathrm{Rank}$ is the rank loss, which can be expressed as
\begin{equation}
    L_\mathrm{Rank} = \sum_{(i,s):C_i<C_s} f(I_s - I_i),
\end{equation}
where $f$ is a hinge loss function $f(x) = \max \{1-x, 0\}$, $I$ is the RPN score, $C$ is the predicted corresponding class logit, the opposite order of $i$ and $s$ under $C$ and $I$ plays a role in sorting. Thus, this pairwise ranking loss can guide the navigator network to find the most helpful order for the classification results. The weights of the losses can be adjusted, but for simplicity and efficiency in hyperparameter tuning, we have set them to be equal as there is no significant difference in their contributions to the model.

\section{Training and Evaluation}\label{sec: training&evaluation}

Our model contains numerous hyperparameters, which can be categorized into three types: those from the model (e.g., number of proposal regions), those from the data (e.g., image size), and those from the training process (e.g., learning rate). We employ different strategies to optimize these parameters. For the model's hyperparameters, the {GPU} memory usage and training time is a crucial limit for number of proposal regions and anchors: only 2--10 proposal regions are possible to fit in the 40 GB memory of an NVIDIA A100, and more anchors require more {CPU} time and CPU--GPU communication overhead. {We also observe a notable decline in performance when experimenting with distributed parallel training across multiple GPUs.} Thus, we tune these parameters within the constraints of available resources. Other models' hyperparameters are also hard to constrain very precisely, such as the number of RPN layers, the convolution kernel sizes, and the number of anchors. These parameters are typically either picked by previous experience \citep{mishkinSystematicEvaluationCNN2017,linFeaturePyramidNetworks2017, yangLearningNavigateFinegrained2018} or tuned through trial and error.

For the data's hyperparameters, they are constrained by both the CNN model and a prior visual check.
{When choosing the input image size for LC and PS, we increase the dimensions along both the time and frequency axes to accommodate the proposed zooming regions.} Since most CNN models are not optimized for large image sizes, we adopt the common sizes (e.g., 112, 128, and their multiples of 2 or 4) following most downstream works of ResNet. {The image sizes used in our model are listed in Table \ref{tab: hyperparameters}.} Another preset prior treatment with some hyperparameters is our image plotting algorithm. We implement algorithms for drawing 1D sequence data to 2D images for LCs and PSs separately: scatter points are mapped in a 2D array directly for LCs, while line connections are drawn with Bresenham's line algorithm \citep{bresenhamAmbiguitiesIncrementalLine1987} for PSs. Without using any visualization package, the pure matrix operation makes our drawing fast. The size for points are 5\,pixels and we use different sizes and gray scales for multiband data.

For the hyperparameters of the training process, {we optimize them using fivefold cross-validation on our training data set}. To save tuning time, we only conduct tests on the variable star data set. Our performance metric is the model accuracy, which is the proportion of instances that are correctly classified. Though we implement the parameter component {of the model}, we do not train it because we want to compare our model with other works that rely solely on LCs.
All models are trained for more than 100 epochs until a significant decrease in the accuracy of the validation set is observed. A learning rate reduction schedule is implemented at the 60-th and 100-th epochs, with each reduction being by 50\%. The maximum number of epochs allowed is set to 500. Various learning rate schedules are tested, but no significant differences are observed.
{We employ automatic mixed precision in our training process, a technique that accelerates deep learning model training by using both single-precision and half-precision formats, thereby saving memory and increasing efficiency.}

The best average training performance with cross-validation for the LC and PS components is shown in Table \ref{tab: cross-validation}. Based on this model, Table \ref{tab: hyperparameters} lists the hyperparameters selected for this work. We generate proposal regions during the training phase and calculate their losses. However, for the output, only the features of the top $N$ proposal regions ($N$ is the number of classification regions) are used. In our experiment, {we select six proposed regions and four classification regions. The numbers of proposed and} classification regions do not have a significant impact on the PS component, but they are more sensitive in the LC component. We choose this value not only based on performance, but also to make it easier to combine the two components by selecting the same number of regions.

Our LC component and PS component can be merged by incorporating an FC layer to fuse their features. Our package includes a parameter component, but for the sake of fair comparison with previous studies, we do not enable it in this experiment. There are two possible approaches for training this combined model. First, the LC and PS components can be trained separately, after which their parameters are fixed, leaving only the final FC layer free for fusion. Alternatively, and this is our preferred method, the combined model can be trained from scratch. While the first approach tends to be simpler and quicker, the latter method generally delivers better performance.

\begin{deluxetable}{cC}
    \tablecaption{Training Performance on Validation Set\label{tab: cross-validation}}
    \tablehead{\colhead{Model Component} & \dcolhead{\rm{Accuracy}}}
    \startdata
    LC Component         & 0.866 \pm 0.008   \\
    PS Component         & 0.931 \pm 0.006
    \enddata
\end{deluxetable}

\begin{deluxetable}{cC}
    \tablecaption{Hyperparameters\label{tab: hyperparameters}}
    \tablehead{\colhead{Hyperparameter} & \dcolhead{\rm{Value}} }
    \startdata
    Batch size & 96 \\
    Learning rate & 5 \times 10^{-4}\\
    Weight decay & 0\\
    Number of proposal regions & 6\\
    Number of classification regions & 4\\
    Input image size & (256, 512)\\
    Zoomed-in LC image size of LC component & (256, 512) \\
    Zoomed-in phase image size of PS component & (224, 224)
    \enddata
    \tablecomments{{The image size is in the format of (height, width).}}
\end{deluxetable}

\subsection{Model Evaluation} \label{sub: evaluation}
{Upon adopting the optimal hyperparameters after hyperparameter tuning, we retrain the model using the complete training data set and subsequently reevaluate its performance on the test data set.} The performance on the test data set is listed in Table \ref{tab: test-performance} for both the variable star data set and the combined data set. The accuracy is the ratio of the number of correctly classified labels to the ground-truth number. Precision, recall, and the $F_1$ score are calculated in a macro-average way:
\begin{equation*}
    \text{Precision} = \frac{\text{TP}}{\text{TP} + \text{FP}},\
    \text{Recall} = \frac{\text{TP}}{\text{TP} + \text{FN}},\
    F_{1} = 2\frac{\text{Precision} \times \text{Recall}}{\text{Precision} + \text{Recall}},
\end{equation*}
where TP, FP, and FN denote true positives, false positives, and false negatives. Generally, our models demonstrate comparable performance across all these metrics, similar to \citet{audenaertTESSDataAsteroseismology2021} for variables, and \citet{sanchez-saezAlertClassificationALeRCE2021} for the combined data set. However, because we mix their data sets to train a more general model, our data set is not as homogeneous as theirs. Therefore, we need to perform separate analyses for different classes of data, which require calculating confusion matrices.

\begin{deluxetable}{ccCCCC}
    \tablecaption{Performances on Test Set\label{tab: test-performance}}
    \tablehead{\colhead{Model Component} & \colhead{Data set} & \dcolhead{\rm{Accuracy}} & \dcolhead{\rm{Precision}} & \dcolhead{\rm{Recall}} & \dcolhead{F_1\  \rm{score}}}
    \startdata
    \multirow{2}{*}{LC Component}    & Variables (Kepler, TESS)     & 0.887 & 0.910 & 0.891 & 0.900   \\
    & Variables + Transients (Kepler, TESS, ZTF{)} & 0.822 & 0.725 & 0.700 &  0.705 \\
    \hline
    \multirow{2}{*}{PS Component} & Variables (Kepler, TESS)     & 0.934 & 0.954 & 0.932 & 0.941 \\
    & Variables + Transients (Kepler, TESS, ZTF) & 0.845  & 0.732 & 0.708 & 0.715 \\
    \hline
    \multirow{2}{*}{Combined}                 & Variables (Kepler, TESS)     &   0.942 & 0.957 & 0.946 & 0.952\\
    & Variables + Transients (Kepler, TESS, ZTF) &   0.870 & 0.762 & 0.779 & 0.757
    \enddata
    \tablecomments{Precision, recall, and $F_1$ score are calculated with macro-averages.}
\end{deluxetable}

\begin{figure}
    \gridline{\fig{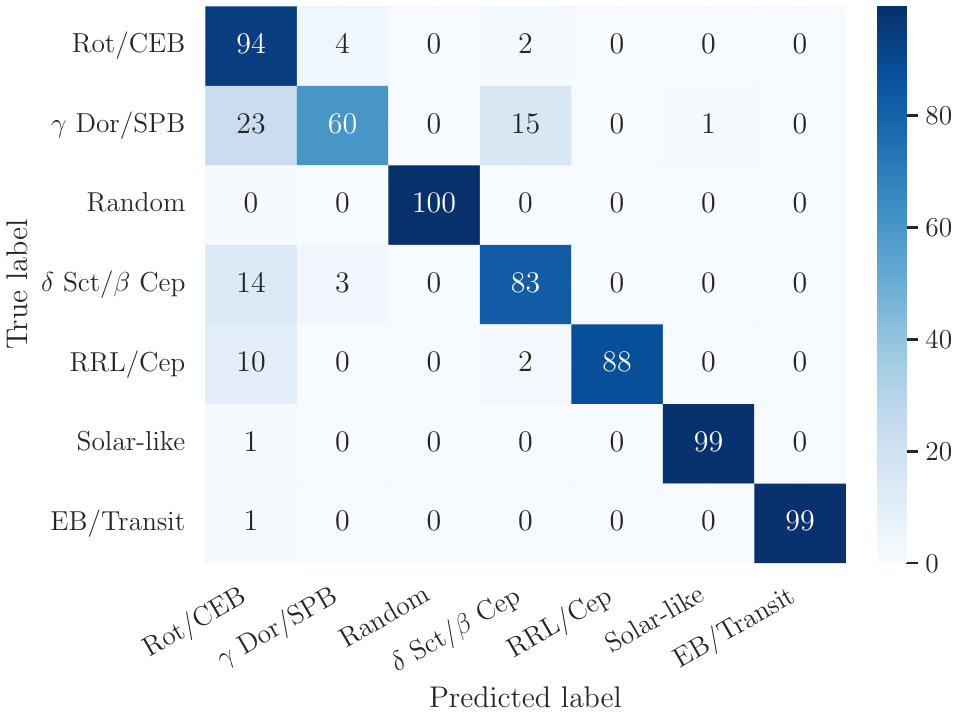}{0.5\textwidth}{(a)}
        \fig{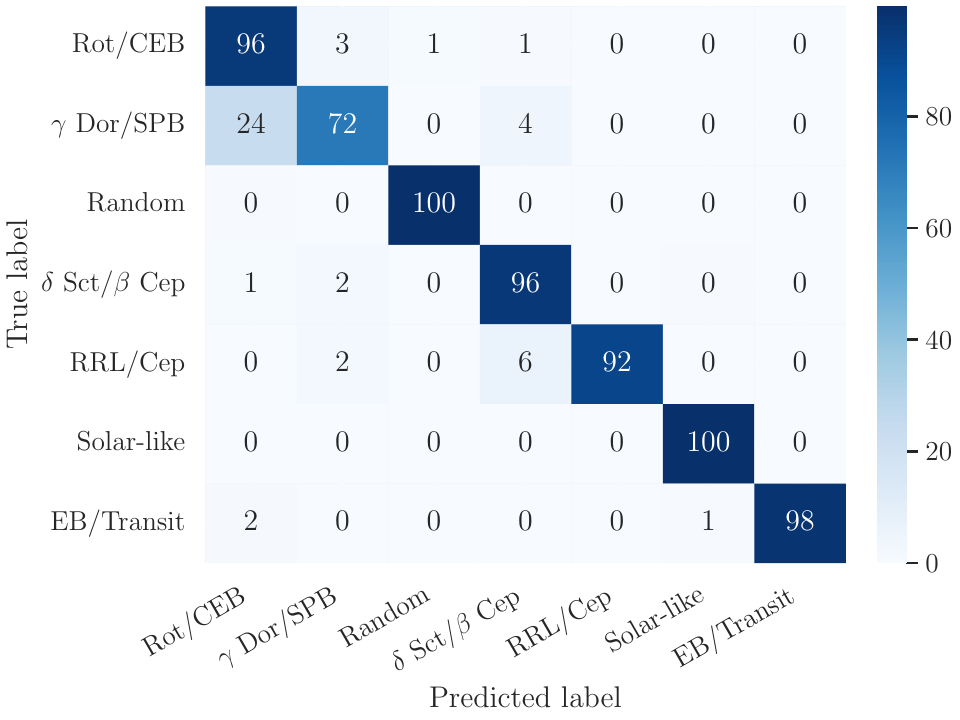}{0.5\textwidth}{(b)}}
    \caption{Normalized confusion matrix of the best LC component (a) and  PS component (b) models on the variable test set {(Kepler, TESS)}. The numbers presented in the matrix represent the proportion of objects that are correctly identified as positive for a specific class (column) out of the total number of objects belonging to that class (row).}\label{fig: var_cm}
\end{figure}

\begin{figure}
    \centering
    \includegraphics[width=0.75\linewidth]{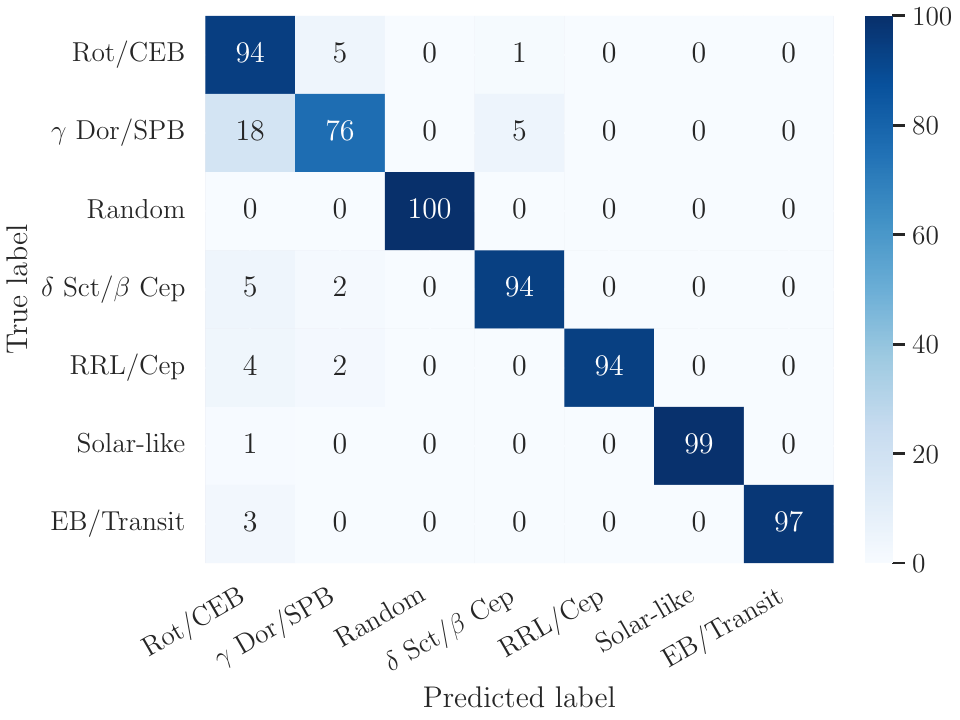}
    \caption{Normalized confusion matrix of the best combined model on the variable test set {(Kepler,  TESS)}. Similar to Figure \ref{fig: var_cm}.}
    \label{fig: comb_var_cm}
\end{figure}

The confusion matrices for the variable star data set are presented in Figure \ref{fig: var_cm} separately for the LC and PS components, and in Figure \ref{fig: comb_var_cm} for the combined model that integrates both LC and PS components. Based on the results presented in Table \ref{tab: test-performance} and Figure \ref{fig: var_cm}, it is evident that the PS component can outperform the LC component, and the combined model returns a better performance than any individual component. This observation aligns with the inherent characteristics of variables. Relative to the T'DA results \citep{audenaertTESSDataAsteroseismology2021}, the accuracy is comparable, but the confusion matrix shows our $\gamma$ Dor/slowly pulsating B stars are mainly misclassified with ratation or contact EBs (Rot/CEBs). This confusion is also somewhat evident in \citet{audenaertTESSDataAsteroseismology2021} because they share a similar parameter space on frequency and variability. However, our relatively weaker performance is mainly due to the TESS samples. The shorter baseline of TESS makes $g$-mode pulsations not as clear as in the Kepler sample in the frequency domain. This confusion can be greatly reduced by introducing stellar physical parameters.

\begin{figure}
    \gridline{\fig{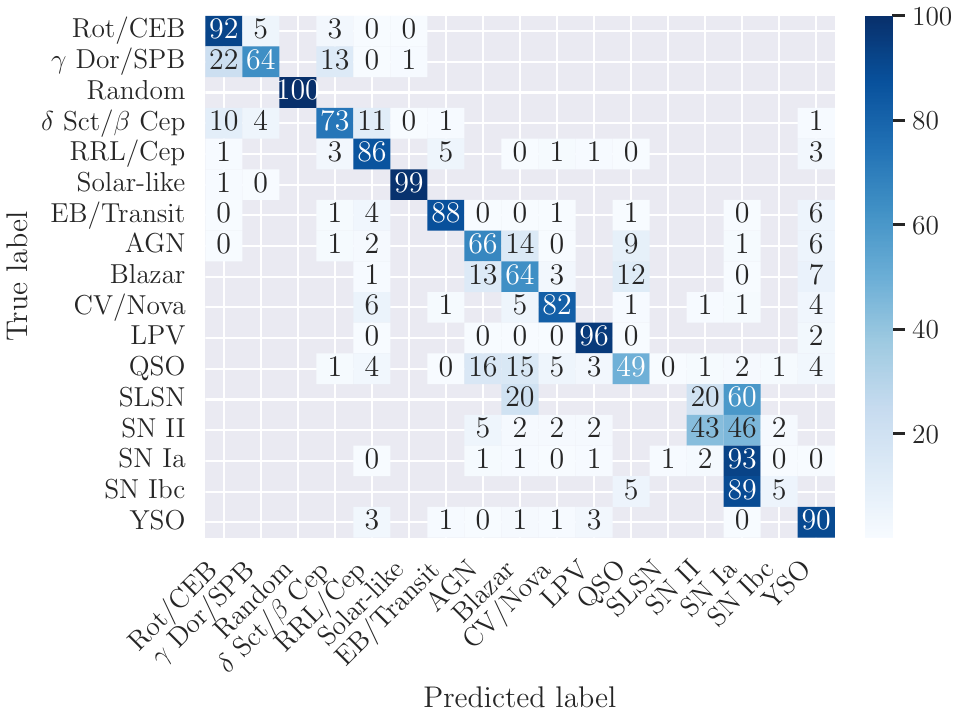}{0.5\textwidth}{(a)}
        \fig{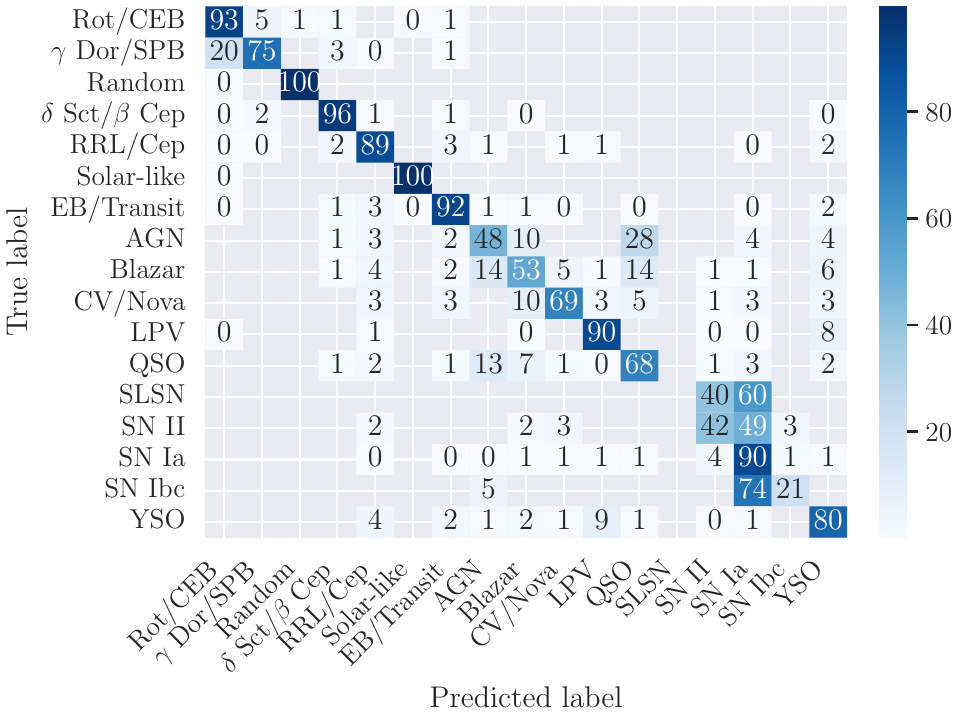}{0.5\textwidth}{(b)}}
    \caption{Normalized confusion matrix of the best LC component (a) and  PS component (b) models on the combined test set {(Kepler, TESS, ZTF)}. Similar to Figure \ref{fig: var_cm}. {The zeros displayed in the figures are rounded numbers. We keep them to ensure correct summation for each row and column.}}\label{fig: comb_cm}
\end{figure}

\begin{figure}
    \centering
    \includegraphics[width=0.75\linewidth]{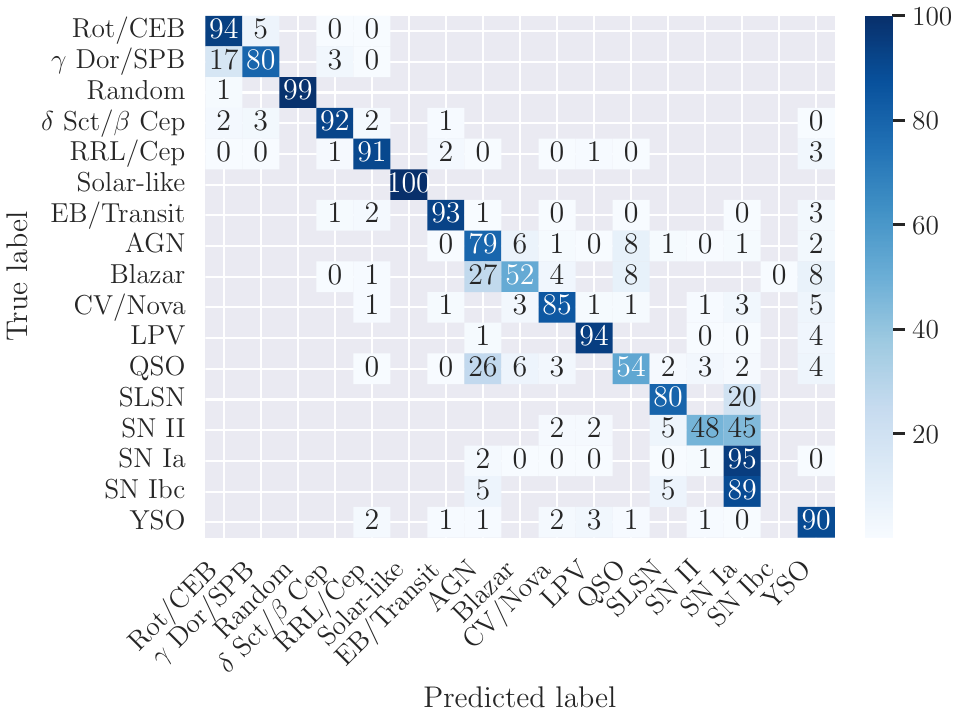}
    \caption{Normalized confusion matrix of the best combined model on the combined test data set {(Kepler, TESS, ZTF)}. Similar to Figure \ref{fig: var_cm}. {The zeros displayed in the figure are rounded numbers. We keep them to ensure correct summation for each row and column.}}
    \label{fig: comb_comb_cm}
\end{figure}

Figures \ref{fig: comb_cm} and \ref{fig: comb_comb_cm} illustrate the confusion matrices of our models of combined data set. As we expect, most transients are more accurately classified in the LC component than in the PS component. This is because their nonperiodic features are clearer in the time domain, while the periodic variables have slightly worse perforamce, especially for the classes containing ZTF LCs. Our models outperform the AleRCE ZTF classification \citep{sanchez-saezAlertClassificationALeRCE2021} in some classes, particularly in the CVs/Novae and young stellar objects. This is due to our direct use of LC morphology information in the LC component, as evidenced by comparison with the PS component. However, our models underperform when it comes to transients, especially SNe Ibc, which are nearly always misidentified as SNe Ia. As explained in \citet{sanchez-saezAlertClassificationALeRCE2021}, this confusion may be intrinsic because the physical mechanisms that cause variability are similar \citep{arnettCribSheetSupernova2008} and cannot be easily distinguished from LC shapes alone. {We also find that the performance on Superluminous SNe is not stable, likely due to their limited sample size.} {Additionally, other transients such as AGN, QSOs, and blazars also exhibit strong confusions; LC shapes are insufficient for distinguishing them. However, redshift and spectral data can alleviate this confusion effectively.} Therefore, more specialized models and additional data (e.g., spectra) are necessary for accurate {transient} classification.

To investigate the impact of zooming ability on performance, we conduct a simple ablation experiment. Specifically, we disable the zooming modules and compute the accuracy of our trained LC component for variable star modeling using only raw LC data. The resulting accuracy is 0.749, which is significantly lower than the 0.887 achieved with zooming enabled (see Table \ref{tab: test-performance}). This suggests that our LC model benefits greatly from zooming functionality. We also apply this experiment to the PS component and find little improvement in accuracy due to most frequency information being captured by the overall PS. However, this does not imply that zooming is unnecessary for the PS component, because it is crucial for model robustness and detecting peculiar targets.

\begin{figure}
    \plottwo{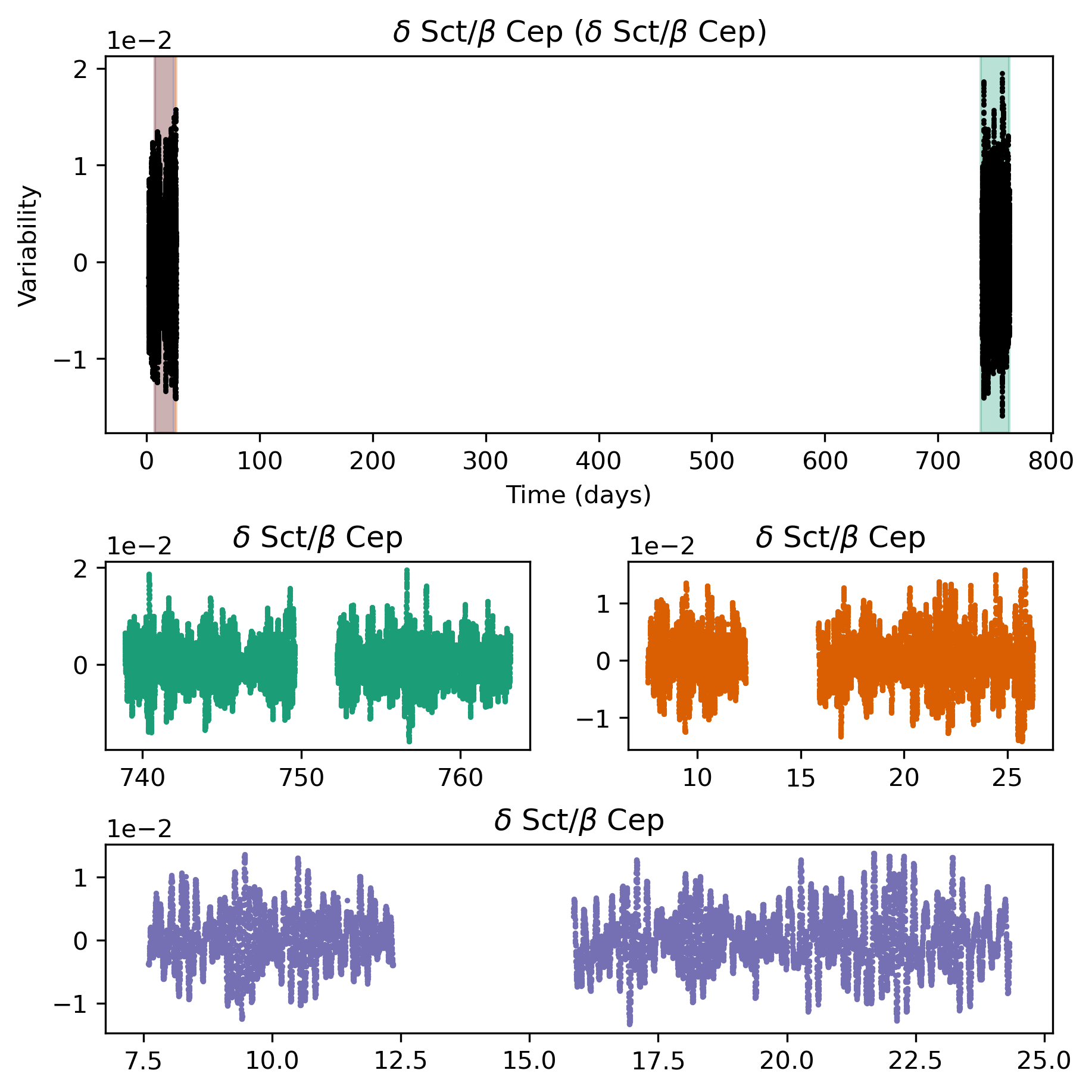}{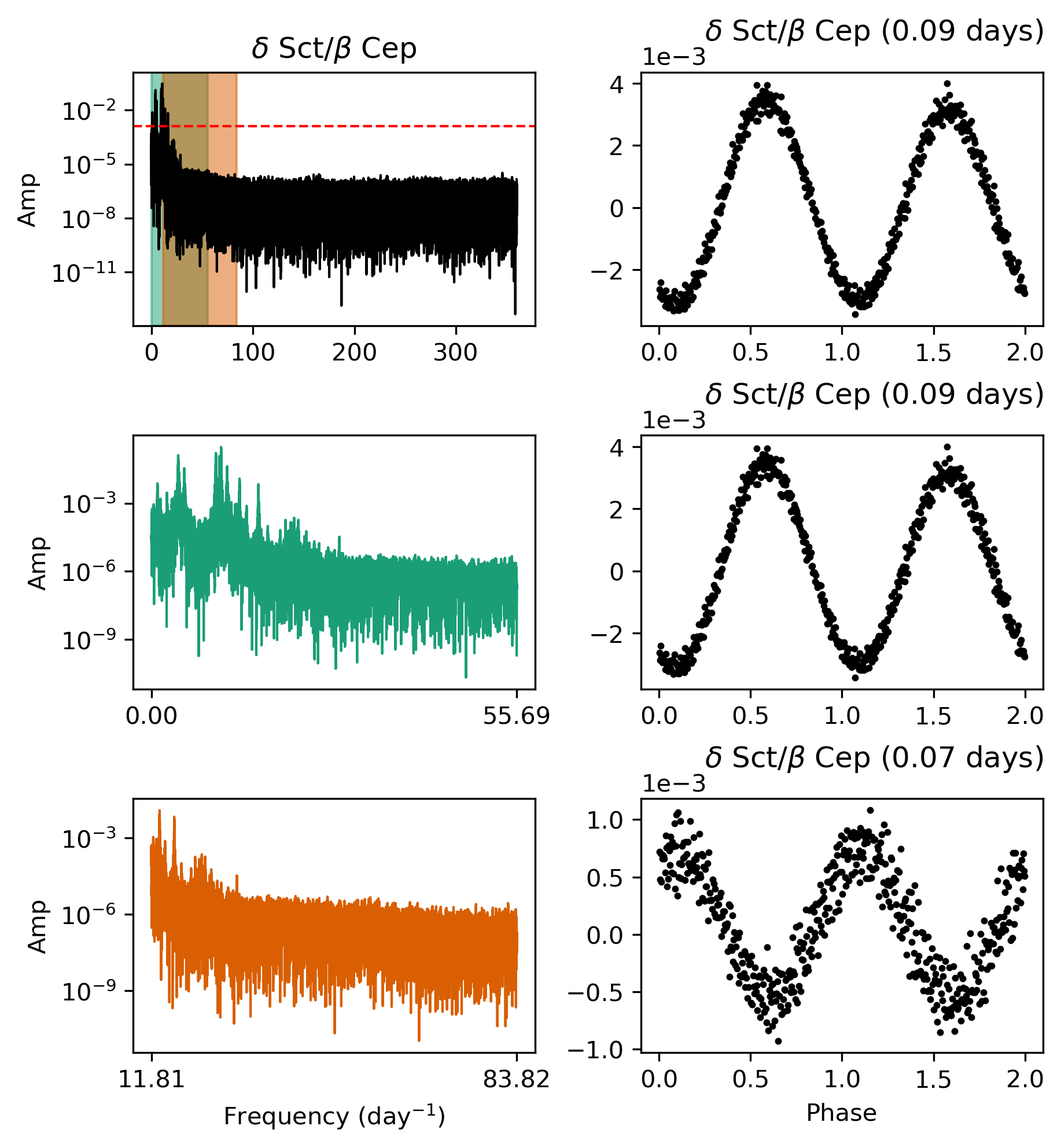}
    \caption{{Classification intermediate results of TIC 397330244, a $\delta$ Scuti star. The left section of the figure shows (top panel) the raw LC with three zoomed-in regions highlighted in semitransparent colors and (middle and bottom panels) the zoomed-in LCs corresponding to the highlighted regions, with predictive labels indicated in their titles. The title of the top panel includes the predicted labels from the results combined with the PS component, as well as those derived solely from the entire raw LC (in brackets).
    On the right, the panels relate to the PS. The left panels depict the PS with zoomed-in regions and a dashed red line indicating the 1\% false-alarm level. The right panels show the corresponding folded LC, based on the most significant peak from the left PS panels. Predictive labels are provided in the titles of the right PS panels, with the top left PS panel title reflecting the label derived from the entire PS component.}}
\label{fig: tess-dsct-example}
\end{figure}

\begin{figure}
    \plottwo{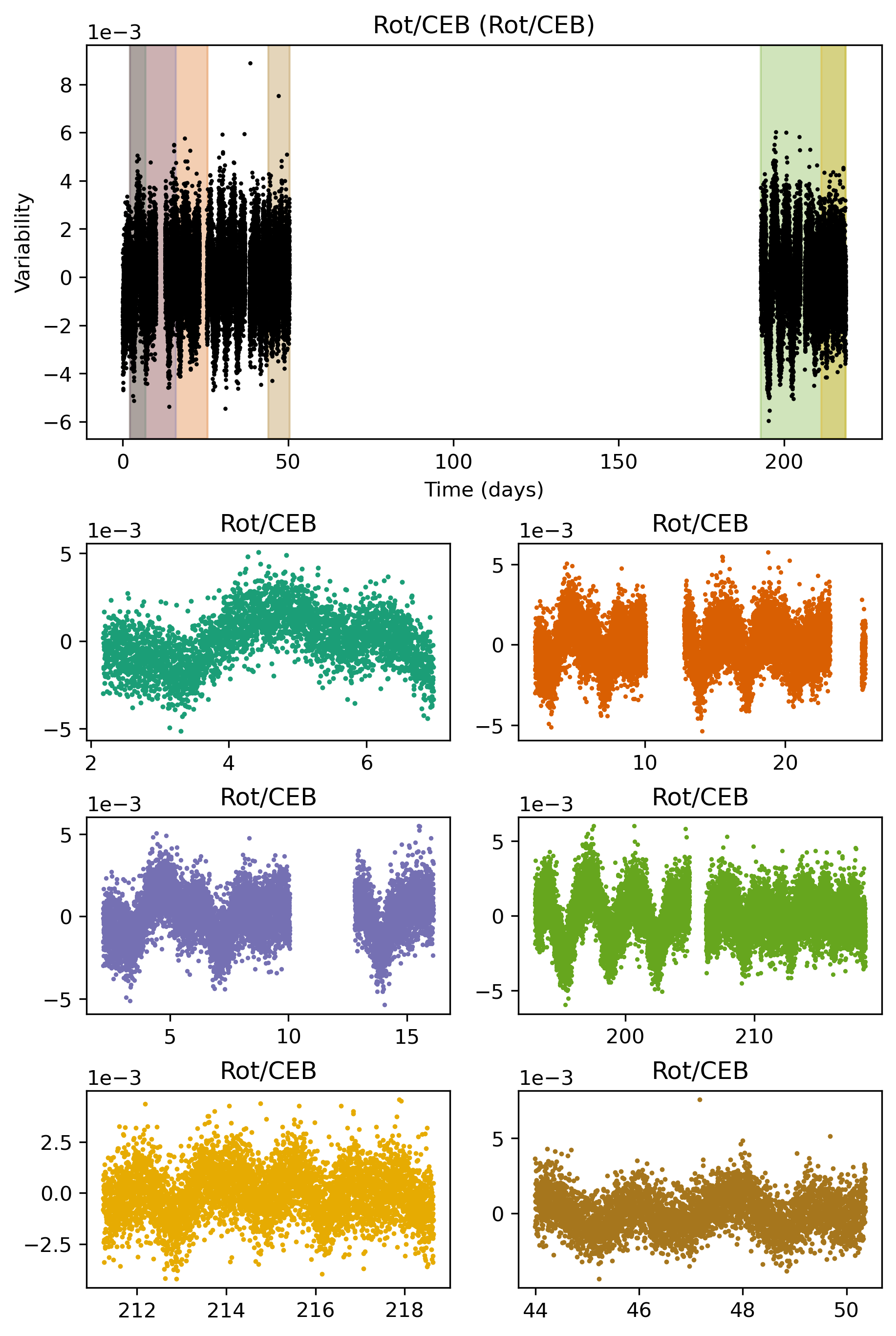}{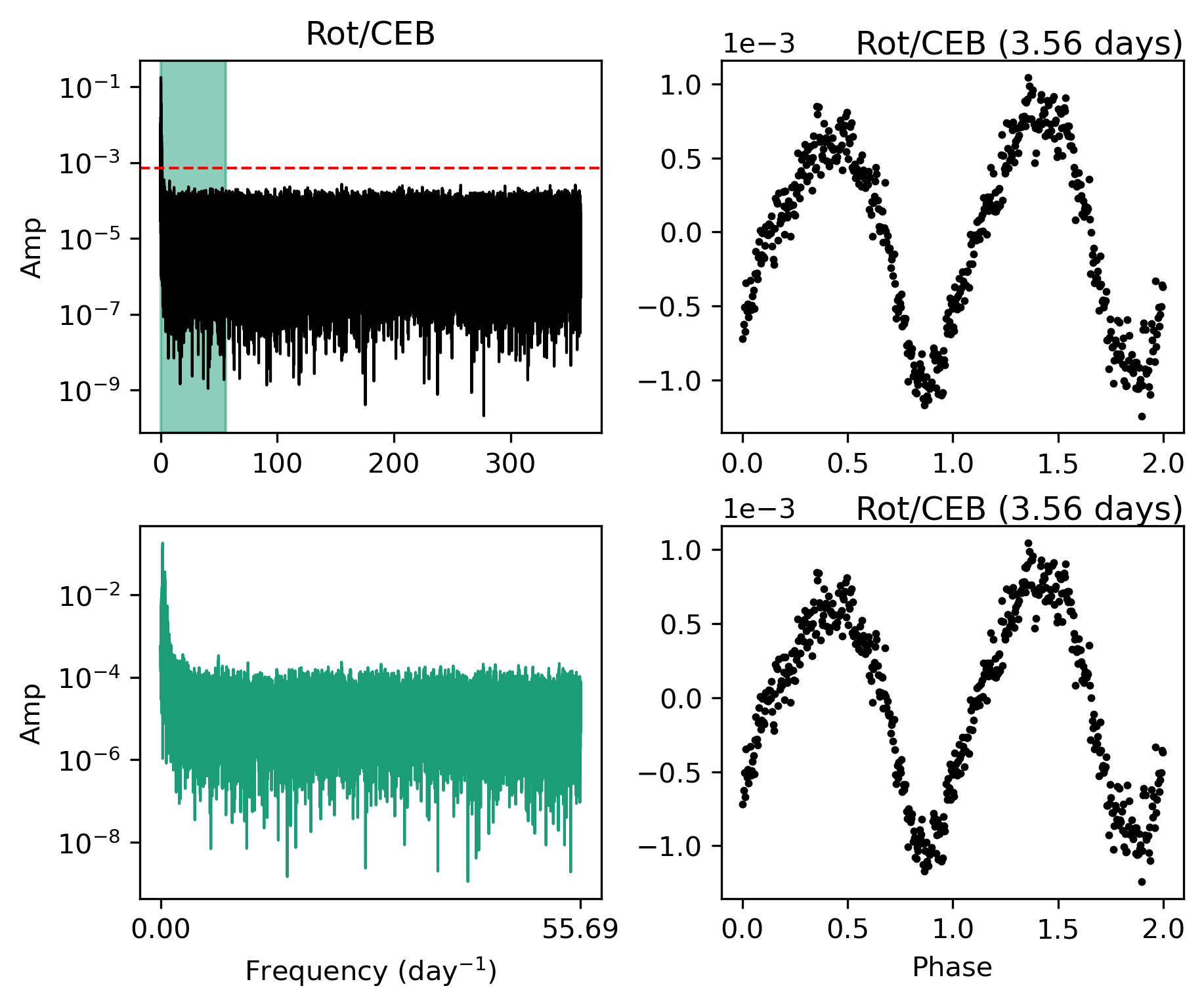}
    \caption{{Intermediate classification results} of TIC 470109695, which is a rotator. Similar to Figure \ref{fig: tess-dsct-example}.}\label{fig: tess-rot-example}
\end{figure}

An important feature of our model is its ability to not only predict class labels, but also provide intermediate results for monitoring the model's performance and conducting further analysis.
To intuitively see the intermediate results of our classification model, we present Figures \ref{fig: tess-dsct-example}--{\ref{fig: tess-rot-example} and Figure \ref{fig: ztf-sn1a-example}--}\ref{fig: kepler-gdor-example}. These figures demonstrate the model's selection of zooming regions and its classification predictions for each region. As shown in Figures \ref{fig: tess-dsct-example}, \ref{fig: tess-rot-example}, and \ref{fig: ztf-eb-example }, our LC component model is capable of accommodating significant gaps in observations by automatically adjusting the zooming windows to focus on different regions. For the observation with only a few points (Figure \ref{fig: ztf-sn1a-example}), the LC component stops zooming in automatically. In many cases, different zoomed-in regions can output different predictions, but the integrated prediction is usually more reliable.

\subsection{Test on Different Data Sets}
\begin{figure}
    \plottwo{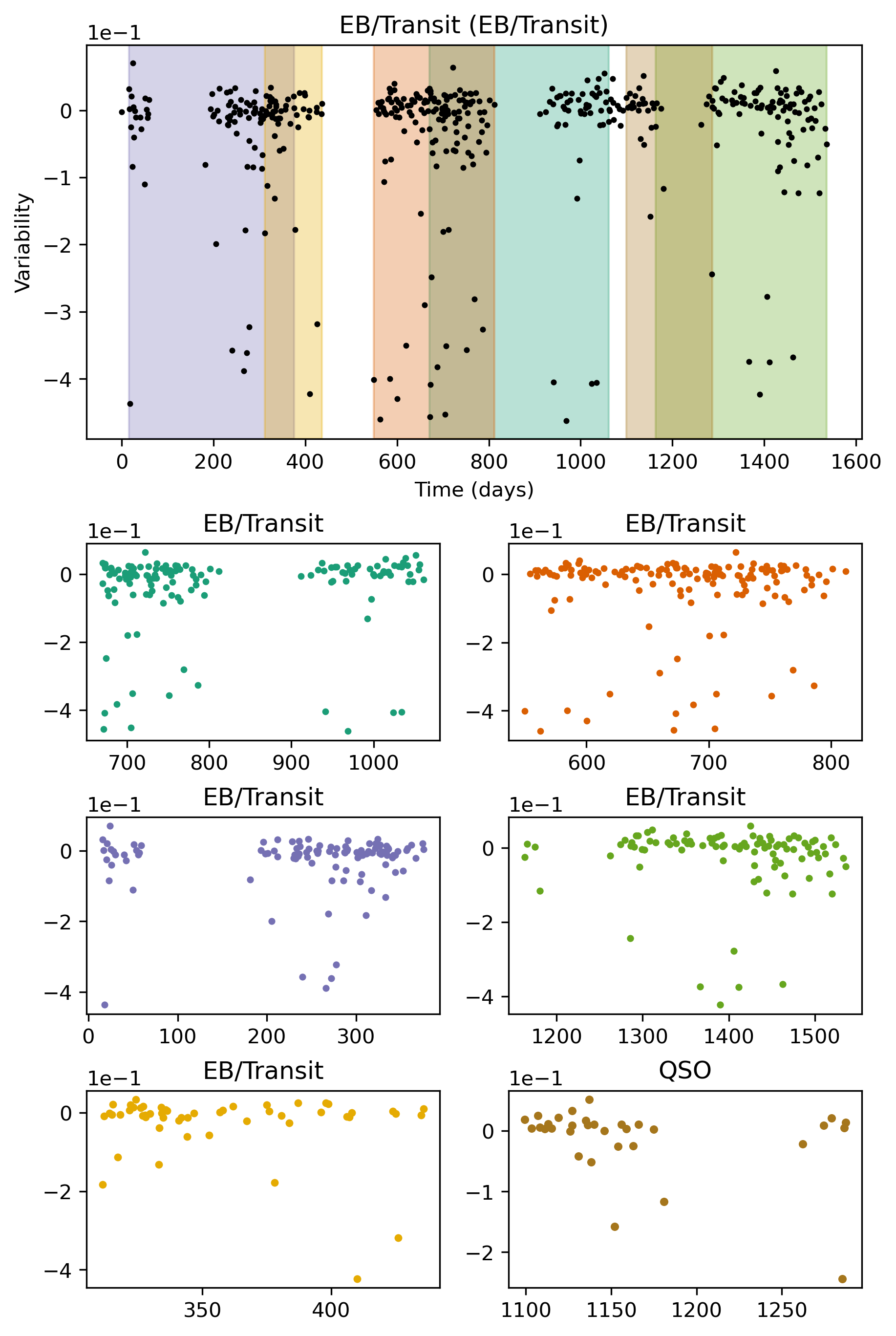}{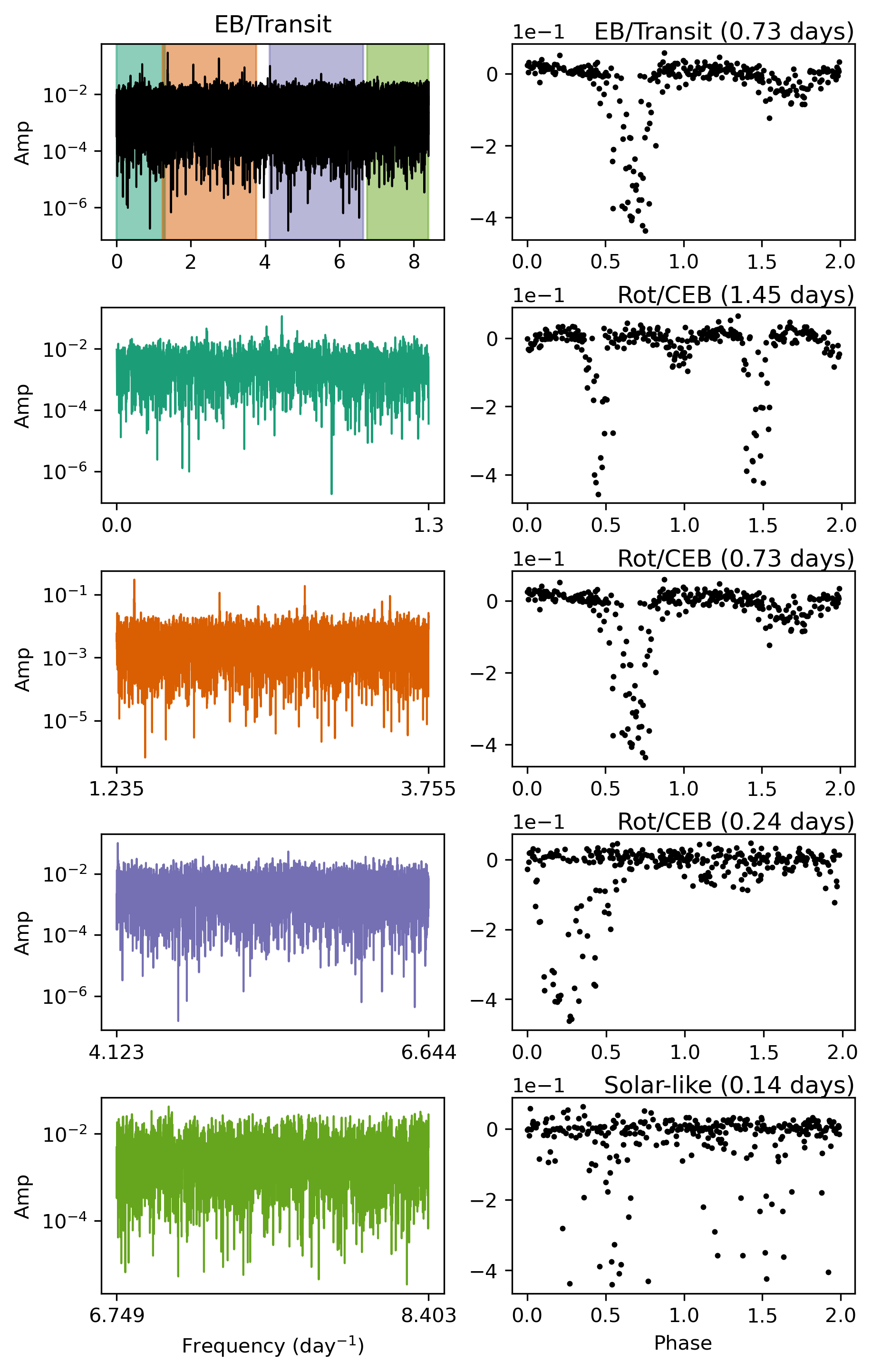}
    \caption{{Intermediate classification results} of ASASSN-V J202436.43+133428.1, which is an EB system correctly classified in our model. Similar to Figure \ref{fig: tess-dsct-example}.}\label{fig: asassn-eb-example}
\end{figure}

\begin{figure}
    \plottwo{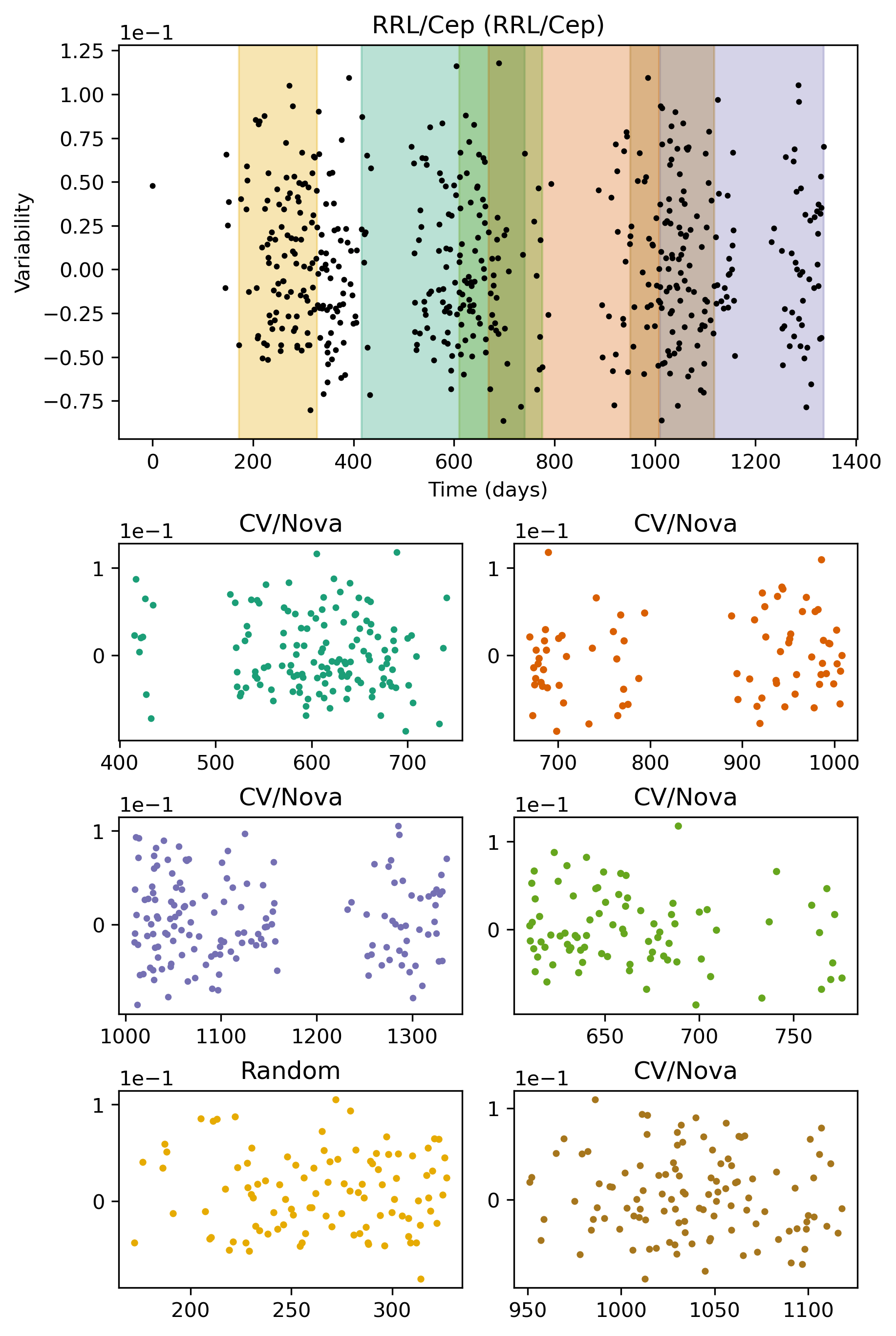}{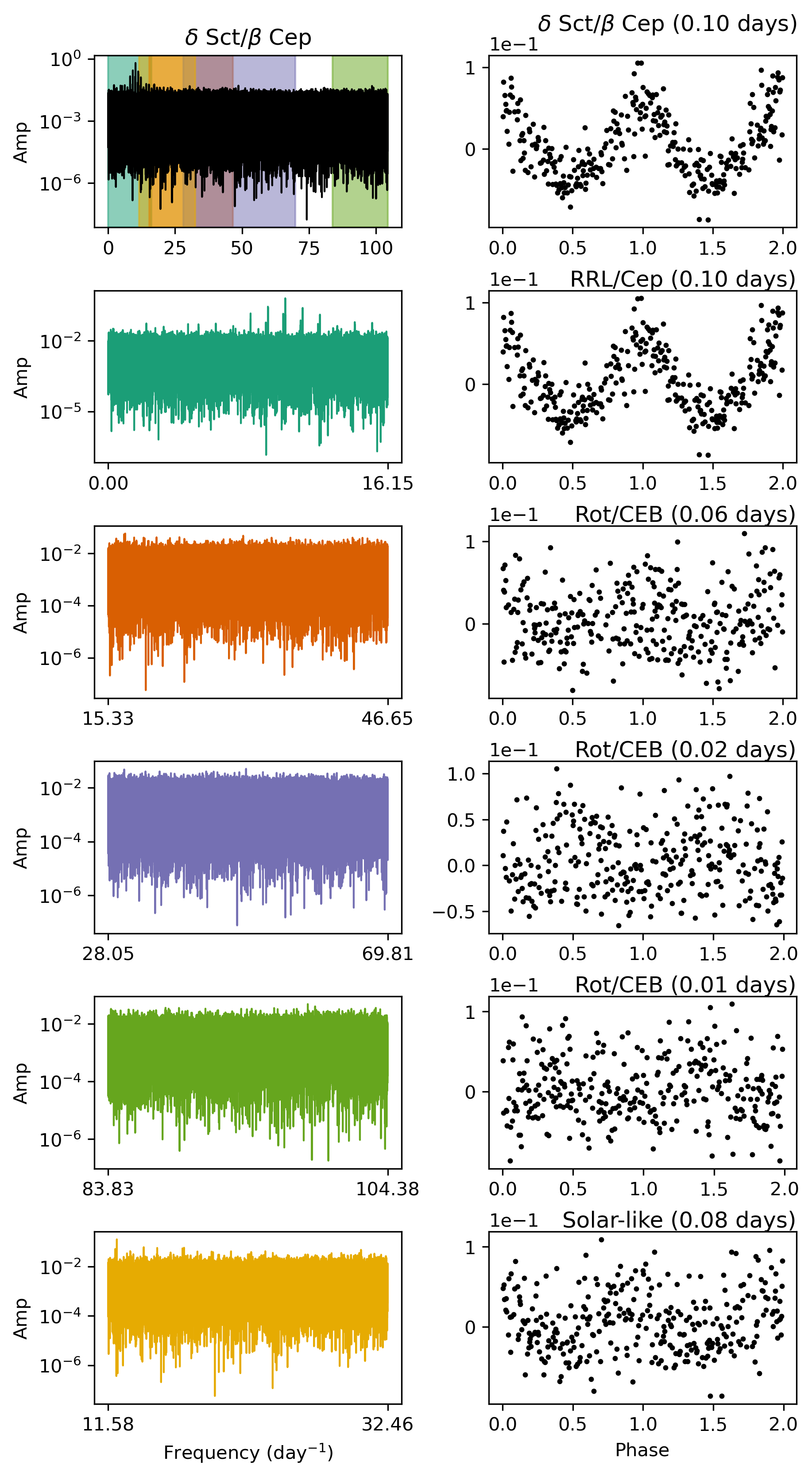}
    \caption{{Intermediate classification results} of ASAS-SN J082703.44-265001.6, which is a $\delta$ Sct object  misclassified by our model. The 95\% conformal predictive set {(see Section \ref{sub: uncertainty})} is [RRL/Cepheid, $\delta$ Sct/$\beta$ Cep, Rot/CEB, EB/transit]. Similar to Figure \ref{fig: tess-dsct-example}.}\label{fig: asassn-dsct-example}
\end{figure}

Since our model is trained on both space- and ground-based data with adaption to different time scales and observational cadences, we can test if it can be general for different missions. Here we choose the ASAS-SN {\it g}-band LCs from the most recent variable catalog \citep{shappeeManCurtainXRays2014,christyASASSNCatalogueVariable2023}. Because ASAS-SN taxonomy is different from ours, {we select the 100 objects with the highest probabilities for each common class. This approach ensures that the chosen objects are more likely to be ground truths, enhancing the reliability of our test results.} Without any fine-tuning and preprocessing on the ASAS-SN LCs, we treat them as single-band LCs and directly send them to our model (trained on combined data set). For ASAS-SN EBs, our model reaches 81\% accuracy; for $\delta$ Sct, we get 37\% accuracy; for RRL, we can get 94\% accuracy. Our model performs with satisfactory accuracy on EBs and RRL for their prominent features, but with quite low accuracy on $\delta$ Sct stars. Figure \ref{fig: asassn-eb-example} displays a correctly classified EB and Figure \ref{fig: asassn-dsct-example} shows a misclassified $\delta$ Scuti star. When comparing with the accurately classified TESS $\delta$ Sct stars (Figure \ref{fig: tess-dsct-example}), it is observed that without physical parameters, a decrease in photometric precision can lead to the misclassification of $\delta$ Sct stars as RRL due to their distorted shape. However, the PS component in Figure \ref{fig: asassn-dsct-example} provides the correct label and indicates an uncertain prediction, which can be utilized {for further analysis (see Section \ref{sub: uncertainty})}.


Dealing with lower photometric precision LCs requires a model that accounts for the uncertainty in the observation, which is not currently included in our model. While we do consider photometric uncertainties when calculating the multiband Lomb--Scargle, other physical quantities and region selections are calculated without accounting for uncertainty. As a result, the performance of our model may decrease when the photometric precision is lower than that of ZTF. To address this issue, more complex data preprocessing or model fitting operations may be required in the future model (see Section \ref{sec: discussion}).

\subsection{Model Uncertainty} \label{sub: uncertainty}
Estimating predictive uncertainty for a machine learning model in a reliable manner is essential for decision making and model interpretability. However, most classical machine learning models and modern deep learning models are observed to have a miscalibration in their confidence \citep{niculescu-mizilPredictingGoodProbabilities2005,guoCalibrationModernNeural2017}.

\begin{figure}
    \centering
    \includegraphics[width=0.6\textwidth]{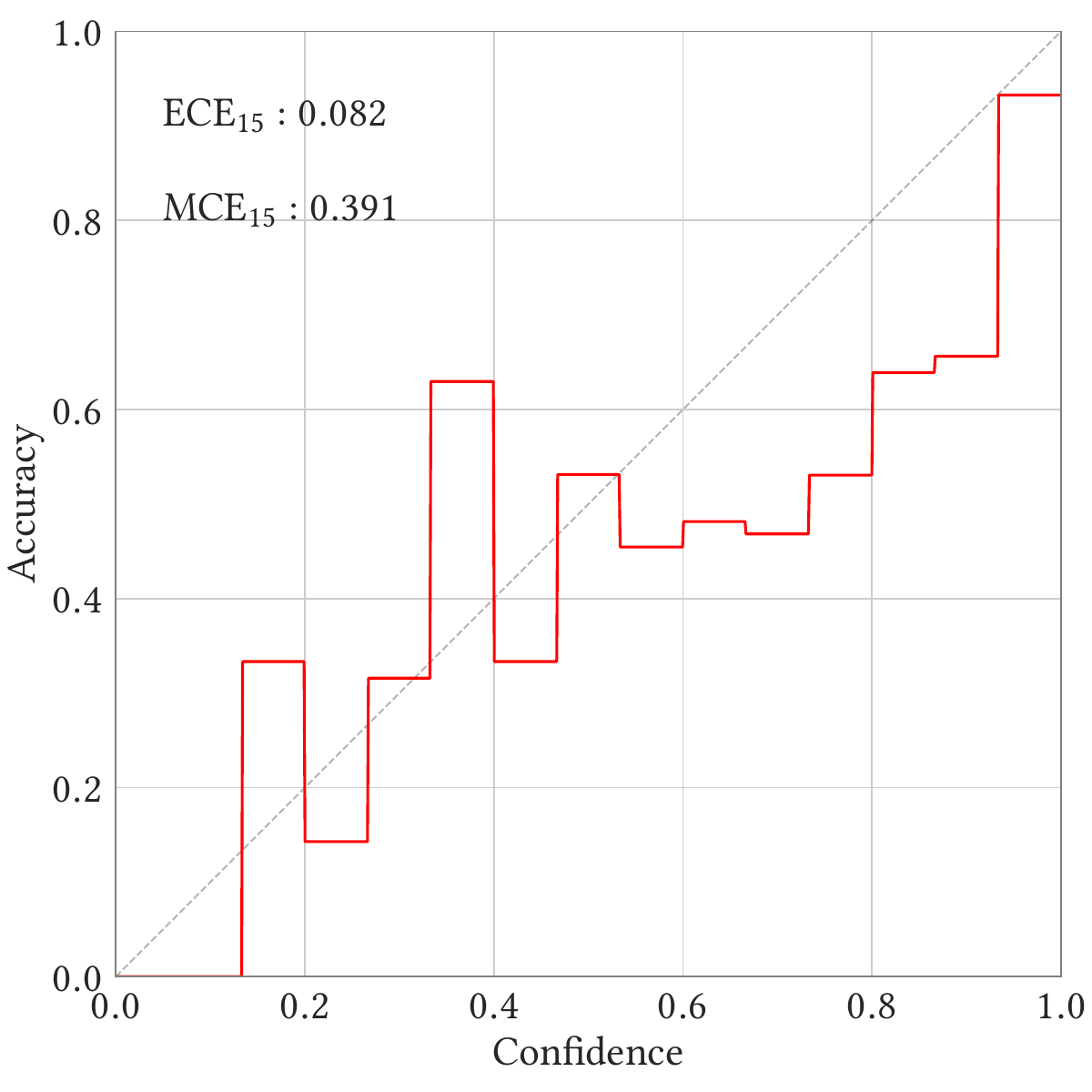}
    \caption{Reliability diagram of our sample. {The x-axis represents the confidence of the model, while the y-axis represents the accuracy of the model. The red step line shows the confidence and corresponding accuracy. The diagonal dashed line represents the ideal calibration, where the confidence is equal to the accuracy. The deviation from the diagonal line indicates the miscalibration of the model.}}
    \label{fig: rel-diagram}
\end{figure}

To address this issue, we calculate the calibration error of our model with the expected calibration error \citep[ECE;][]{naeiniObtainingWellCalibrated2015} and maximum calibration error \citep[MCE;][]{naeiniObtainingWellCalibrated2015}. ECE is the expectation of difference between confidence and accuracy, which can be estimated as
\begin{align}\label{eq: ece_mce}
    \text{ECE} & = \sum_i^N b_i |p_i - c_i|, \\
    \text{MCE} & =  \max_{i} (p_i - c_i),
\end{align}
where $N$ is the number of bins, and for each bin $i$, $p_i$ is the prediction accuracy, and $c_i$ is the average confidence score. The parameter $b_i$ is the fraction of data points in bin $i$. An ideal calibrated classifier would have both the ECE and MCE equal to 0.
In this study, we select a commonly used value of $N = 15$ and obtain the ECE of 0.082 and the MCE of 0.391 for our optimal combined model trained on the combined data set. These results are consistent with previous studies conducted on various CNN models \citep[ECE 4\%--16\%, MCE ~20\%--40\%][]{guoCalibrationModernNeural2017}, indicating common miscalibration. Figure \ref{fig: rel-diagram} shows the corresponding reliability diagram, which is the expected sample accuracy as a function of confidence \citep{guoCalibrationModernNeural2017}, and the deviation from the diagonal line indicates imperfect calibration.

In order to quantify the uncertainty of our deep learning classifier and make our model more applicable to decision making, we employ a technique based on conformal prediction \citep{vovkMachineLearningApplicationsAlgorithmic1999}, a framework that can produce prediction sets with provable coverage guarantees. It guarantees that the prediction sets are designed to contain the true label of a new test instance with a high probability, regardless of the underlying data distribution or model assumptions.
{Specifically, we adopt the Regularized Adaptive Prediction Sets \citep[RAPS;][]{angelopoulosUncertaintySetsImage2022} algorithm to calculate our final predictive sets. The RAPS algorithm can adjust our classifier to produce predictive sets that include the true label with a user-defined probability.}

To calibrate our model, we {create} a calibration data set {by splitting} our test data set {into two equal parts: a calibration data set and a validation data set}. After {calibrating} our best-performance model with RAPS at the 5\% significance level (95\% coverage), we show the calibration performance in Table \ref{tab: raps-performance}, where the accuracy is calculated for the validation data set. {The Top-1 accuracy is calculated by considering only the prediction with the highest probability within the predictive set, while the Top-5 accuracy takes into account the predictions of the top five probabilities. Since the validation data set is derived from the test data set, the Top-1 accuracy is very similar to the accuracy results in Table \ref{tab: test-performance}.} The predictive set coverage is closer to 95\%, consistent with the prespecified 5\% confidence level. {The size of the prediction sets is an important indicator of the model's uncertainty. A smaller size indicates a more confident prediction, while a larger size indicates a more uncertain prediction.} The average sizes of the prediction sets are also listed in Table \ref{tab: raps-performance}. {Our combined model has a slightly larger average size for the data set containing both variables and transients. This also indicates higher uncertainty within the combined data set.}

\begin{table}
    \caption{Conformal calibration results with 95\% coverage rate}\label{tab: raps-performance}
    \begin{tabular}{ccclcc}
        \hline\hline
        \multirow{2}{*}{Model Component} & \multirow{2}{*}{Data Set}                   & \multicolumn{2}{c}{Accuracy} & \multirow{2}{*}{Coverage} & \multirow{2}{*}{Size}        \\
        \cline{3-4}
                                         &                                            & \multicolumn{1}{l}{Top-1}    & Top-5                                                    \\
        \hline
        \multirow{2}{*}{Combined}        & Variables (Kepler, TESS)                   & 0.934                        & 0.999                     & 0.935                 & 1.11 \\
                                         & Variables + Transients (Kepler, TESS, ZTF) & 0.877                        & 0.991                     & 0.953                 & 1.79 \\
        \hline
    \end{tabular}
\end{table}

After conformal calibration, our model becomes more applicable for unseen samples. For instance, when we test our conformally calibrated model to ASAS-SN $\delta$ Sct objects as discussed in Section \ref{sec: training&evaluation}, the 95\% coverage predictive sets return an accuracy of 79\%, which is significantly better than the point prediction. However, the average size of these prediction sets is 3.99, which is larger than our calibrated average value. This excessive length of the prediction sets reflects the model's uncertainty and may serve as an indicator for anomalous targets. We will further elaborate on this in Section \ref{sec: peculiar detection}.

\section{Peculiar Object Detection} \label{sec: peculiar detection}
The goal of typical classification algorithms is to achieve accurate and unambiguous predictions. However, in practice, there are often objects that are difficult to classify or that do not fit into any known category. These objects may be known types but are not included in the training data set, or they could be something currently unknown. This issue is crucial yet difficult and is generally referred to as out-of-distribution (OOD) or out-of-sample detection. Related problems such as anomaly detection, novelty detection, and outlier detection are the same in essence but with varying degrees of unseen samples.

Once a model is able to distinguish whether a sample comes from in-distribution (ID) or OOD origins, it can be more general and reliable for practical astronomical observations. In addition to classification, it may also provide direct assistance in the search for unknown peculiar targets.
In recent astronomical observations, there has been a notable increase in detecting LC anomalies due to the growing demand for the identification of uncommon transients \citep[e.g.,][]{pruzhinskayaAnomalyDetectionOpen2019,malanchevAnomalyDetectionZwicky2021,villarDeeplearningApproachLive2021}, where they applied an isolation forest algorithm in the context of active learning. Inspired by KIC 8462852 \citep[Boyajian’s star;][]{boyajianPlanetHuntersIX2016}, some more general LC anomaly detection has also been performed on Kepler data \citep[][]{gilesDensitybasedOutlierScoring2020,martinez-galarzaMethodFindingAnomalous2021}.
However, their anomalies are detected mainly based on the clustering of features from the data. Instead of using the raw features from the data, since we already have a well-performing classifier, we can abstract classification features along more physically meaningful directions to apply OOD detection.
To do so, we apply a few different approaches at different levels.

\begin{figure}
    \centering
    \plottwo{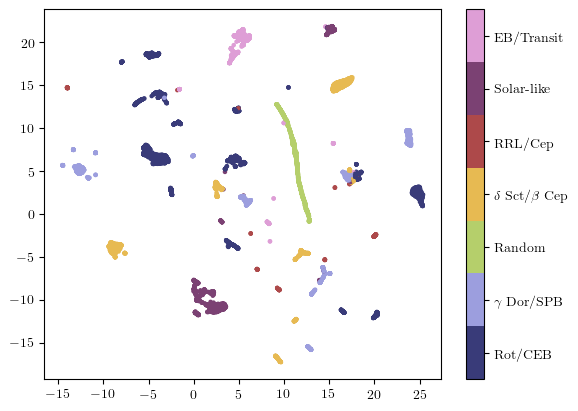}{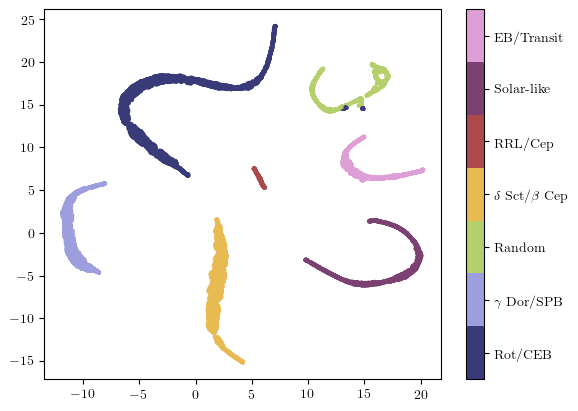}
    \caption{{UMAP projections of the penultimate feature layers of the LC and PS components.} The left {panel} displays the UMAP of the LC component, while the right {panel} shows the UMAP of the PS component. The different colors represent different classes of objects. 
    }
    \label{fig: umap}
\end{figure}

Traditionally, unsupervised dimensionality reduction techniques algorithms such as t-SNE \citep{maatenVisualizingDataUsing2008} and uniform manifold
approximation and projection \citep[UMAP;][]{mcinnesUMAPUniformManifold2020} are often used for {visualizing high-dimensional data in two or three dimensions, which aids in} outlier detection. Figure \ref{fig: umap} displays the UMAP {projection results} of the penultimate feature layer, i.e., the layer before the FC layer. The LC component trained on a data set consisting of variables shows some distinct groups corresponding to our predefined classes. Compared {to} the UMAP of the PS component (right subfigure of Figure \ref{fig: umap}), the LC component's features show multiple clusters at different locations, indicating worse performance of OOD detection on the LC component. However, these algorithms are primarily designed for visualization and exploration of high-dimensional data, they have some inherent issues such as sensitivity to hyperparameters, and the lack of built-in outlier probability measures may hinder their efficacy. Here we do not use them for OOD detection but present them as an intuitive visualization of the penultimate features.

Recent progress in OOD detection on image tasks makes full use of the features of neural networks and employs various methods to attempt to obtain OOD results \citep{yangGeneralizedOutofDistributionDetection2022}. We implement some of these methods and apply them in our LC classification framework. At first, to evaluate the performance of those OOD metrics, we have to create an OOD sample.
However, creating an OOD sample is difficult because the concept of OOD can be at different levels for different scenarios. In this study, we select heartbeat stars from Kepler \citep{prsaKEPLERECLIPSINGBINARY2011,abdul-masihKeplerEclipsingBinary2016}\footnote{\url{http://keplerebs.villanova.edu/}} as the OOD sample. {Heartbeat stars are a type of binary star system with close periastron passage in a highly eccentric orbit. We choose this type of star because it is not included in our training data set, and it has a unique LC morphology compared to other variables.}

A lot of OOD metrics are available. For instance, employing the
maximum softmax probability \citep[MSP;][]{hendrycksBaselineDetectingMisclassified2018} score is the most straightforward way to distinguish between ID and OOD data. The
Out-of-distribution Detector for Neural Networks \citep[ODIN;][]{liangEnhancingReliabilityOutofdistribution2020} introduces temperature scaling and input preprocessing so that ID data and OOD data have more distinct distributions.
The energy score \citep{liuEnergybasedOutofdistributionDetection2021} is a modification of logits borrowing the expression of free energy and the partition function.
Virtual-outlier synthesis \citep[VOS;][]{duVOSLearningWhat2022} generates virtual outliers on the penultimate layer based on the multivariate Gaussian distribution conditioned by the classes. Directed
sparisification \citep[DICE;][]{sunDICELeveragingSparsification2022} prunes the less contributed neurons to make the OOD data more separate from ID data. Virtual-logit matching \citep[ViM;][]{wangViMOutOfDistributionVirtuallogit2022} calculates an additional logit representing the virtual OOD class generated from the residual of the feature against the principal space, and then matches it with the original logits by a constant scaling. The probability of this virtual logit after softmax is the indicator of OOD-ness.

\begin{figure}
    \centering
    \includegraphics[scale=0.8]{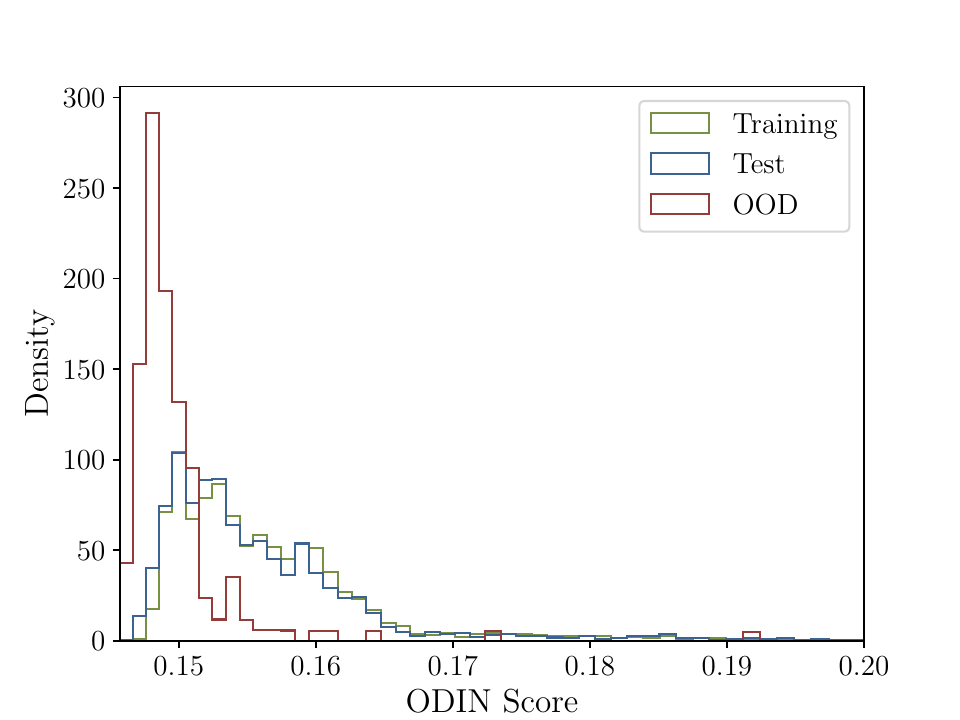}
    \caption{ODIN score distribution {of various data sets {with only the PS component enabled}. The legend indicates that the different colored lines represent ODIN score distributions derived from distinct data sets. Both the training and test data sets are sourced from the variable data set, whereas the OOD data set is obtained from heartbeat stars.} The distribution is truncated for ODIN scores {above} 0.2.}
    \label{fig:ood-odin}
\end{figure}

\begin{figure}
    \centering
    \includegraphics[scale=0.8]{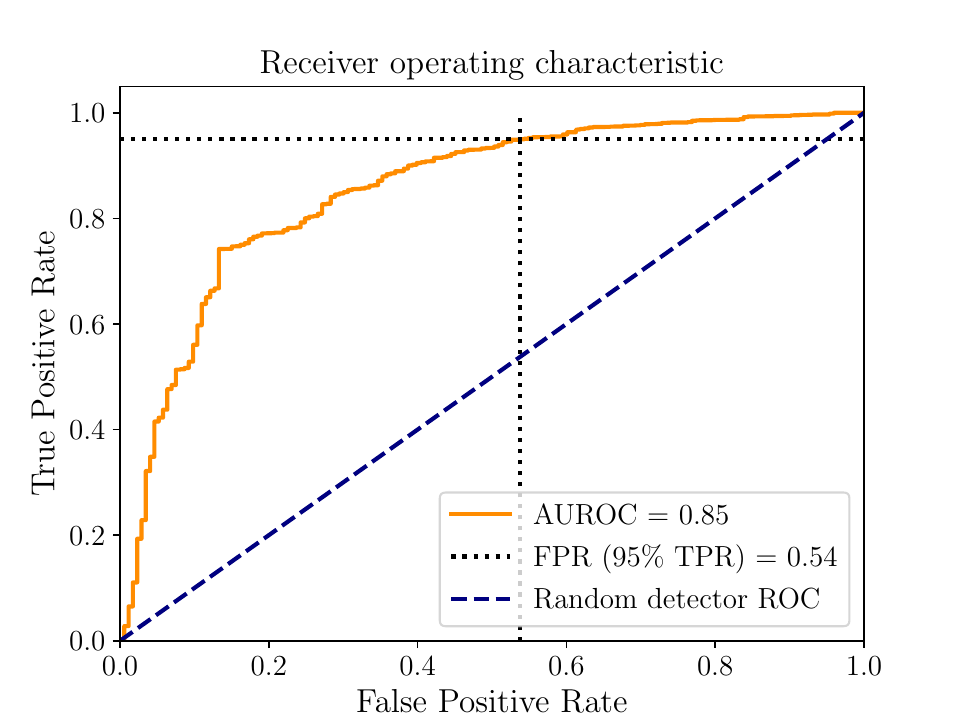}
    \caption{{ODIN ROC curve {with only PS component enabled}, represented by the orange line, is generated by varying the ODIN score threshold. The x-axis denotes the FPR, and the y-axis denotes the TPR. The diagonal dashed blue line illustrates the performance of a random detector. The dotted black lines indicate the 95\% TPR and corresponding FPR.}}
    \label{fig:ood-odin-roc}
\end{figure}

Among the metrics mentioned above, VOS needs to retrain the model, while DICE requires replacement of the last FC layer, and ViM aligns the projection with logits through the weights and biases from the FC layer. Consequently, we retrain our model, enabling VOS in the process. Following this, we compute the DICE and other metrics, using both the VOS-enabled and VOS-disabled models for comparison. The best OOD detection performance comes from the PS component trained on variable stars. {In OOD detection, similar to many other binary classification tasks, performance is typically evaluated using the receiver operating characteristic (ROC) curve since the metric score itself is task-specific and not particularly meaningful. In the ROC curve, by varying the OOD metric scores, the true-positive rate (TPR) for OOD objects changes in relation to the false-positive rate (FPR) for ID objects. Therefore, the threshold score for distinguishing OOD objects can be directly obtained from the ROC curve by setting a target TPR.} We utilize the area under the ROC (AUROC) and FPR at 95\% TPR (FPR@TPR95) as metrics to evaluate the efficiency of OOD detection. A higher AUROC indicates better ability to distinguish between OOD and ID, while a lower FPR@TPR95 signifies fewer misclassifications of ID samples as OOD when 95\% of OOD samples are correctly detected. {Table \ref{ood-table} lists the OOD performance with different OOD metrics.} Figure \ref{fig:ood-odin} displays the ODIN score distributions {of} the training, testing, and OOD data sets. It is evident that the training and test samples have identical distributions, but OOD data is located in a lower ODIN score region and can be {partially} distinguished from training/test data using a threshold. This threshold can be determined with the ROC curve shown in Figure \ref{fig:ood-odin-roc}. For a TPR of 95\%, the threshold is 0.149; lower TPR values result in higher thresholds.
Retraining models from scratch for VOS affects model performance on ID data. In our case, we achieve an accuracy of 0.928 for PS components on variable data sets, a result that is acceptable in most instances.

We also calculate OOD metrics for the LC component and our combined model. For the LC component, the best AUROC is 0.65, which is significantly worse than that of the PS component {(0.85)}. This suggests that either the metrics used are not sensitive enough to detect OOD samples, or as shown in the UMAP plot, the features from the LC component are much more complex than those from the PS component and cannot be easily represented by simple OOD indicators.
Our combined models have three FC layers after the penultimate feature layer, and thus VOS, DICE, and ViM are not available; the best AUROC is just slightly above 0.5, indicating that the combined model may be suffering from the same issue as the LC component and can hardly distinguish the OOD sample with current metrics. {While our OOD metrics show good performance in identifying heartbeat stars, there are many potential types of OOD objects that may yield varying results across different metrics. Consequently, we cannot ensure consistent performance for all possible OOD scenarios.}


\begin{deluxetable}{ccccc}
    \tablecaption{OOD Performace of PS component\label{ood-table}}
    \tablehead{\multicolumn{3}{c}{OOD Detection Algorithms} & \dcolhead{\rm{AUROC}\,\uparrow} & \dcolhead{\rm{FPR@TPR95}\,\downarrow}}
    \startdata
    \multirow{8}{*}{With VOS}    & \multirow{4}{*}{With DICE}    & MSP     &  0.80     &    0.88       \\
    &                               & ODIN    &  0.80     &    0.89       \\
    &                               & Energy &  0.84     &    0.71       \\
    &                               & ViM     &  0.73     &    0.90       \\
    \cline{2-5}
    & \multirow{4}{*}{Without DICE} & MSP     &  0.76     &    0.77       \\
    &                               & ODIN    &  \bf{0.85}    &    \bf{0.54}      \\
    &                               & Energy &  0.53     &    1.00       \\
    &                               & ViM     &  0.60     &    0.94       \\
    \cline{1-5}
    \multirow{8}{*}{Without VOS} & \multirow{4}{*}{With DICE}    & MSP     &   0.69    &    0.88       \\
    &                               & ODIN    &   0.72    &    0.81       \\
    &                               & Energy &   0.63    &    0.98       \\
    &                               & ViM     &   0.83    &    0.63       \\
    \cline{2-5}
    & \multirow{4}{*}{Without DICE} & MSP     &   0.67    &     0.94      \\
    &                               & ODIN    &   0.68    &     0.94      \\
    &                               & Energy &   0.70    &     0.95      \\
    &                               & ViM     &   0.80    &     0.79
    \enddata
    \tablecomments{{The upward arrow indicates that a higher score corresponds to better performance, while the downward arrow signifies that a lower score is preferable. The best performance is highlighted in bold.}}
\end{deluxetable}

Detecting OOD objects is a challenging task.  The reason machine learning classifiers can fail in real-world tasks is that there are discrepancies between the training and testing distributions. These classifiers frequently yield high-confidence predictions that are, unfortunately, incorrect. If the model does not have the capability to flag potential errors, its adoption could be restricted, or it could even lead to severe accidents. A well-calibrated model can be helpful when encountering OOD samples. As we discussed in Section \ref{sub: uncertainty}, our conformally calibrated model can return a predictive set with guaranteed coverage at a specific probability, the predictive sets give a hint to return the OOD objects. For example, we apply our conformally calibrated model to the known Kepler oscillating EB sample \citep{gaulmeSystematicSearchStellar2019}, obtaining an accuracy of 0.896. For the correctly classified objects, 47.6\% of them have predictive sets with multiple candidates, {suggesting} possible pulsating components.

\section{Discussion}\label{sec: discussion}
While our model is general for different time lengths and cadences, it does have its limitations. For instance, it cannot effectively handle extremely short cadence data such as gravitational waves, which possess a sampling rate of 16\,kHz. Processing this type of data in the time domain is wholly unsuitable. Furthermore, extremely short cadences may exceed our frequency calculation range. As the frequency increases, so does the computation time, and this could be problematic.
Lengthy time-series like combining two independent missions together with a large observation gap are possible in our model, but unless the characteristics are notably distinguishable, the model's zooming ability may fail. Potential solutions can include increasing the image size or iteratively applying the model to the selected regions.
Currently, our model only supports two colors. Additional colors can be seamlessly implemented as different gray scales or appended as new image channels. If the input requires many colors (more than 20), the development of new methods may be required. 
Moreover, the impact of systematics on our model can be substantial, affecting both the time and frequency domains. As systematics are influenced by the unique characteristics of different surveys, there is no universal solution for them. Our current model is trained using preprocessed data, specifically the PDCSAP from Kepler/TESS and corrected ZTF LCs. Therefore, the input LC for our model must be at least corrected for systematic signals.

\section{Conclusion}\label{sec: conclusion}
\subsection{Summary}
{In this paper, we provide a general LC classification framework. This framework contains three components, namely, the LC component, PS component, and parameter component. The LC component utilizes a CNN with zooming capabilities to directly process raw LCs and their images, yielding classification results. Similarly, the PS component employs CNN to analyze Lomb--Scargle periodograms and their corresponding images for classification purposes. The parameter component is an FC neural network designed to process physical parameters and generate classification results. The final classification result is derived by integrating the outputs from all available components.}

{Our model is trained on a combined data set from Kepler, TESS, and ZTF. It achieves an accuracy of 0.94, precision of 0.96, recall of 0.95, and an $F_1$ score of 0.95 for the classification of variable stars in macro-averaged metrics. For the classification of transient stars, it attains an accuracy of 0.87, precision of 0.76, recall of 0.78, and an $F_1$ score of 0.76 in macro-averaged metrics.}

{We also apply our trained model directly to the ASAS-SN data set. Despite the varying systematics and photometric precision across different missions, our model achieves satisfactory results for certain types of variable stars, such as EBs and RRL stars, without any retraining or fine-tuning. We further calibrate our model using conformal prediction, which generates robust predictive sets rather than simple labels. By evaluating various OOD detection algorithms, we identify the ODIN metric as having the best OOD performance for variables. All of our models are available,\footnote{\url{https://zenodo.org/records/10081600}} allowing users to select models based on their specific needs.}

\subsection{Final Remarks and Perspective}
Beyond normal optical LCs, our model can be applied to general time-series classifications, such as thoes at other wavelengths like radio, X-ray, or any narrowband long-term observations. However, for these specific cases, the model would require retraining from scratch.
In this work, we deliberately limit our data set to facilitate comparison with previous performance; more complete data sets from various sources are available for practical use.

{Our classification framework will be expanded to include more detailed categories and support multiple labels for a single target. By utilizing the parameter component, we will optimize our model using several other established missions, ongoing surveys such as the Tianyu Project \citep{fengTianyuSearchSecond2024} and Sitian Project \citep{liuSiTianProject2021}, and simulated data sets from the Vera Rubin Observatory's PLAsTiCC project \citep{kesslerModelsSimulationsPhotometric2019}. Additionally, since our model can be applied to most photometric systems, our classifier provides a model that can be directly used or easily fine-tuned for various small-scale observation projects.}

Current large language models demonstrate superior proficiency in understanding natural language. They can replace or extend our parameter component while accepting our intermediate data along with some incomplete and unstructured data (e.g., instrument-specific systematics diagnostics, external environment information, and literature conclusions).
Future possible advanced machine learning models with multimodal capabilities may have the ability to reason and imagine at a human level. These networks can autonomously assess the completeness of information and seek additional data, including experimental manipulation of raw data (e.g., fitting data with the model and adjusting parameters for comparison). This could lead to highly reliable and interpretable classification and analysis models in the future.

\begin{acknowledgments}
    {We appreciate the referee's suggestions and comments, which greatly enhanced the readability of this work.}
    We would like to express our gratitude to Masataka Aizawa for his valuable suggestions and to Shuai Zha and Zhenyu Zhu for generously providing computing resources. We would also like to extend our thanks to L.A. Balona for supplying his complete catalog. We acknowledge the
    support from the B-type Strategic Priority Program of the Chinese Academy of Sciences, grant No. XDB41000000, and the China Postdoctoral Science Foundation (No. 2022M712083). The collaboration of this work is supported by the SJTU Global Strategic Partnership Fund (2022 SJTU-Warwick). This work is also partly supported by the Shanghai Jiao Tong University 2030 Initiative. The authors acknowledge the National Natural Science Foundation of China (General Program Grant No. 12473066 and Key Program Grant No. 11933004). The authors acknowledge the Siyuan-1 cluster supported by the Center for High Performance Computing at Shanghai Jiao Tong University, and the Tsung-Dao Lee Institute AstroCluster for providing GPU resources that have contributed to the research results reported within this paper. Part of the languages have been polished with the assistance of Microsoft Azure OpenAI Service \citep{microsoftcorporationAzureOpenAIService2023}, utilizing both the OpenAI GPT-3.5-0301 \citep{ouyangTrainingLanguageModels2022} and GPT-4-0613 \citep{openaiGPT4TechnicalReport2023} models.
\end{acknowledgments}

%

\vspace{5mm}
\facilities{Kepler, TESS, ZTF}


\software{
    Astropy \citep{theastropycollaborationAstropyCommunityPython2013,theastropycollaborationAstropyProjectBuilding2018,theastropycollaborationAstropyProjectSustaining2022},
    Numpy \citep{harrisArrayProgrammingNumPy2020},
    PyTorch \citep{paszkePyTorchImperativeStyle2019},
    Matplotlib \citep{hunterMatplotlib2DGraphics2007},
    Jupyter Notebook \citep{kluyverJupyterNotebooksPublishing2016},
    Lightkurve \citep{lightkurvecollaborationLightkurveKeplerTESS2018,geert_barentsen_2021_4603214},
    Deep-LC \citep{cuiDeepLCV012024},
    Lightkurve-ext \citep{cuiLightkurveextExtensionLightkurve2024},
    Microsoft Azure OpenAI Service \citep{microsoftcorporationAzureOpenAIService2023}
}



\appendix
\restartappendixnumbering
\section{Examples of Classification Results} \label{sec: classification examples}

\begin{figure}
    \centering
    \begin{minipage}{0.49\textwidth}
        \centering
        \includegraphics[width=\textwidth]{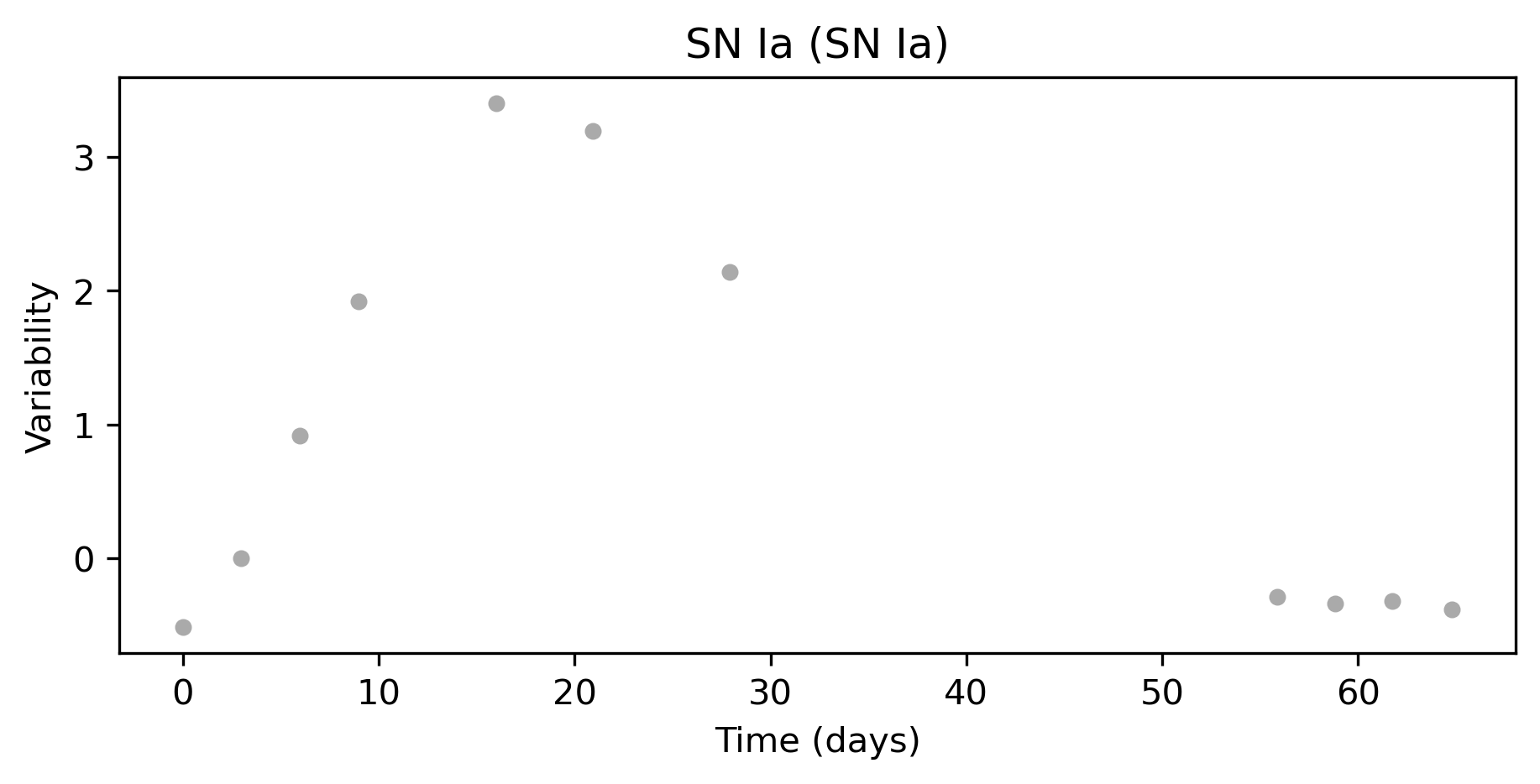}
        \includegraphics[width=\textwidth]{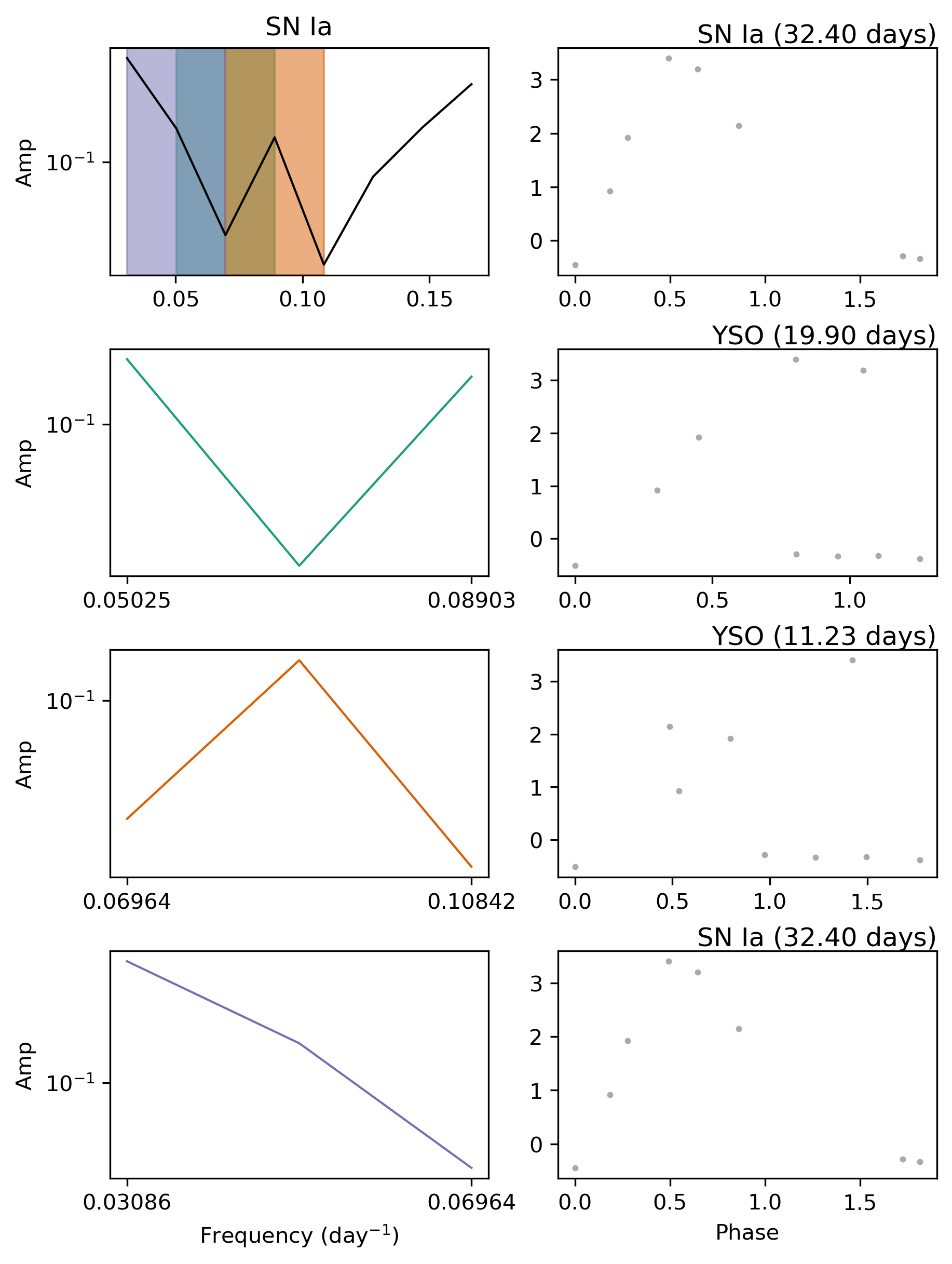}
        \caption{{Intermediate classification results} of ZTF 18abuhyjv, which is an SN Ia from ZTF alert system. Similar to Figure \ref{fig: tess-dsct-example} but plotted as gray scale for data from different observational bands. {The gray points indicate the {\it g} band.}}\label{fig: ztf-sn1a-example}
    \end{minipage}\hfill
    \begin{minipage}{0.49\textwidth}
        \centering
        \includegraphics[width=\textwidth]{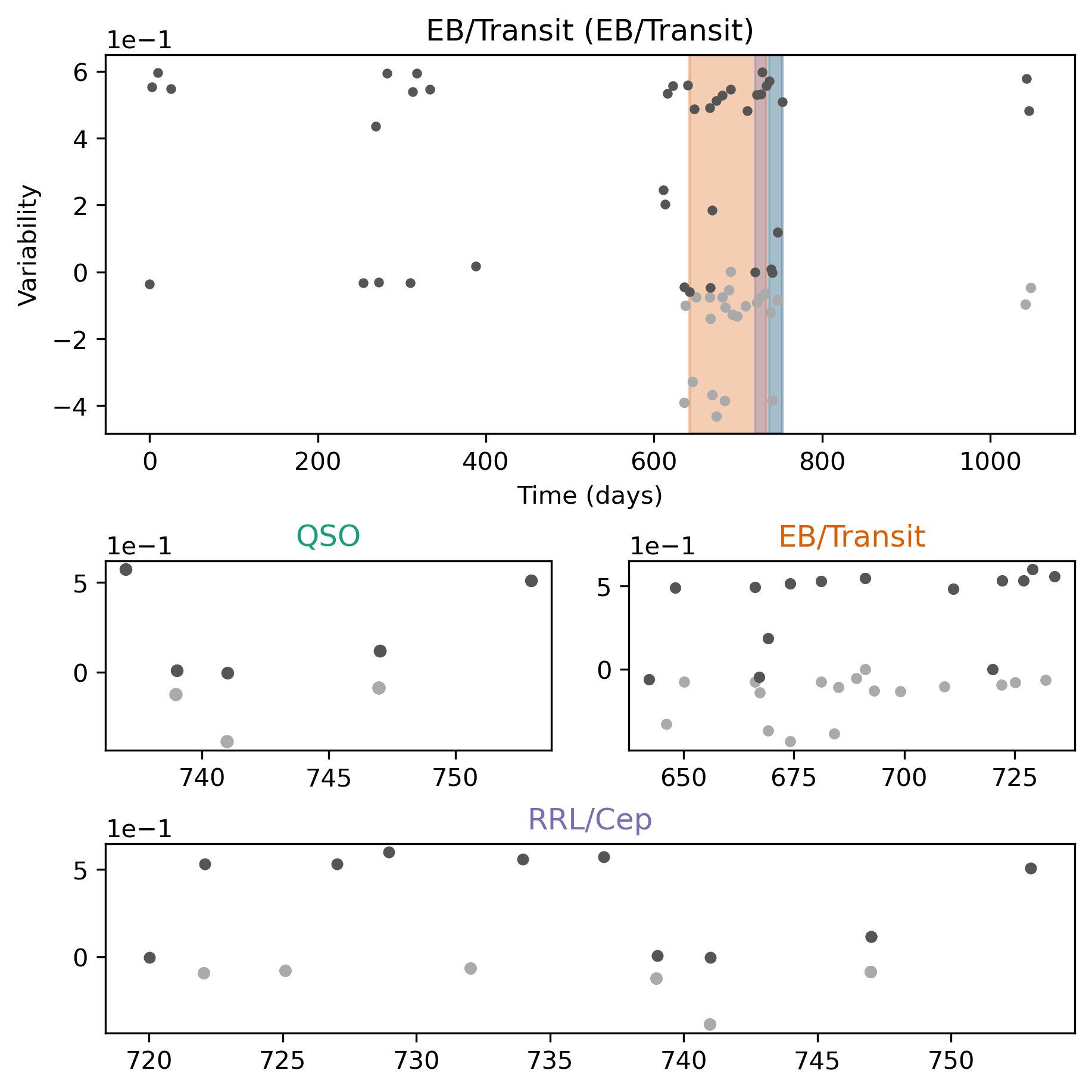}
        \includegraphics[width=\textwidth]{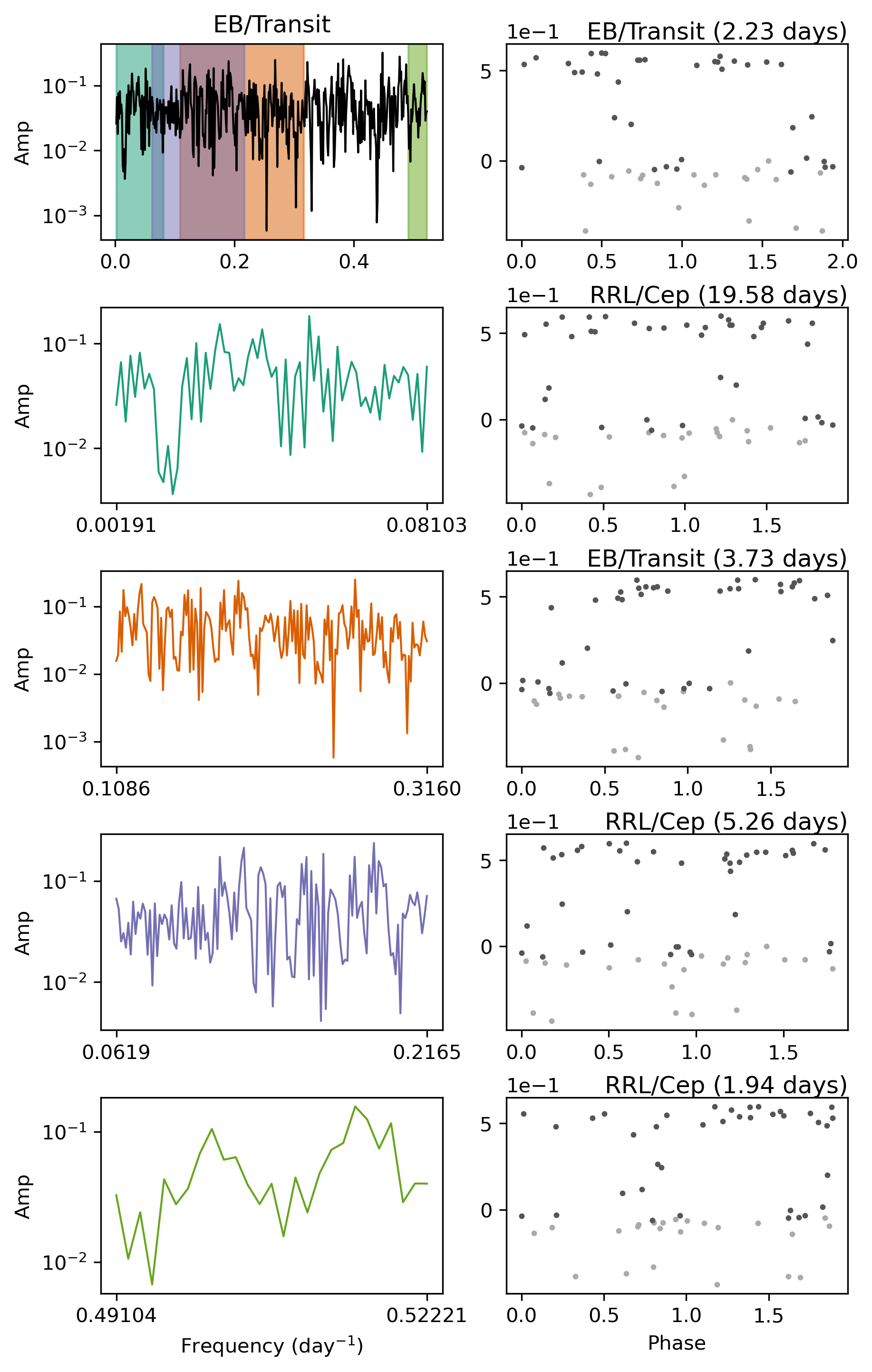}
        \caption{{Intermediate classification results} for ZTF 19aapkqyk, an EB identified by the ZTF alert system. Similar to Figure \ref{fig: ztf-sn1a-example}{; black points represent the {\it r} band, while gray points denote the {\it g} band.}}\label{fig: ztf-eb-example }
    \end{minipage}
\end{figure}

\begin{figure}
    \centering
    \begin{minipage}{0.49\textwidth}
        \centering
        \includegraphics[width=0.9\textwidth]{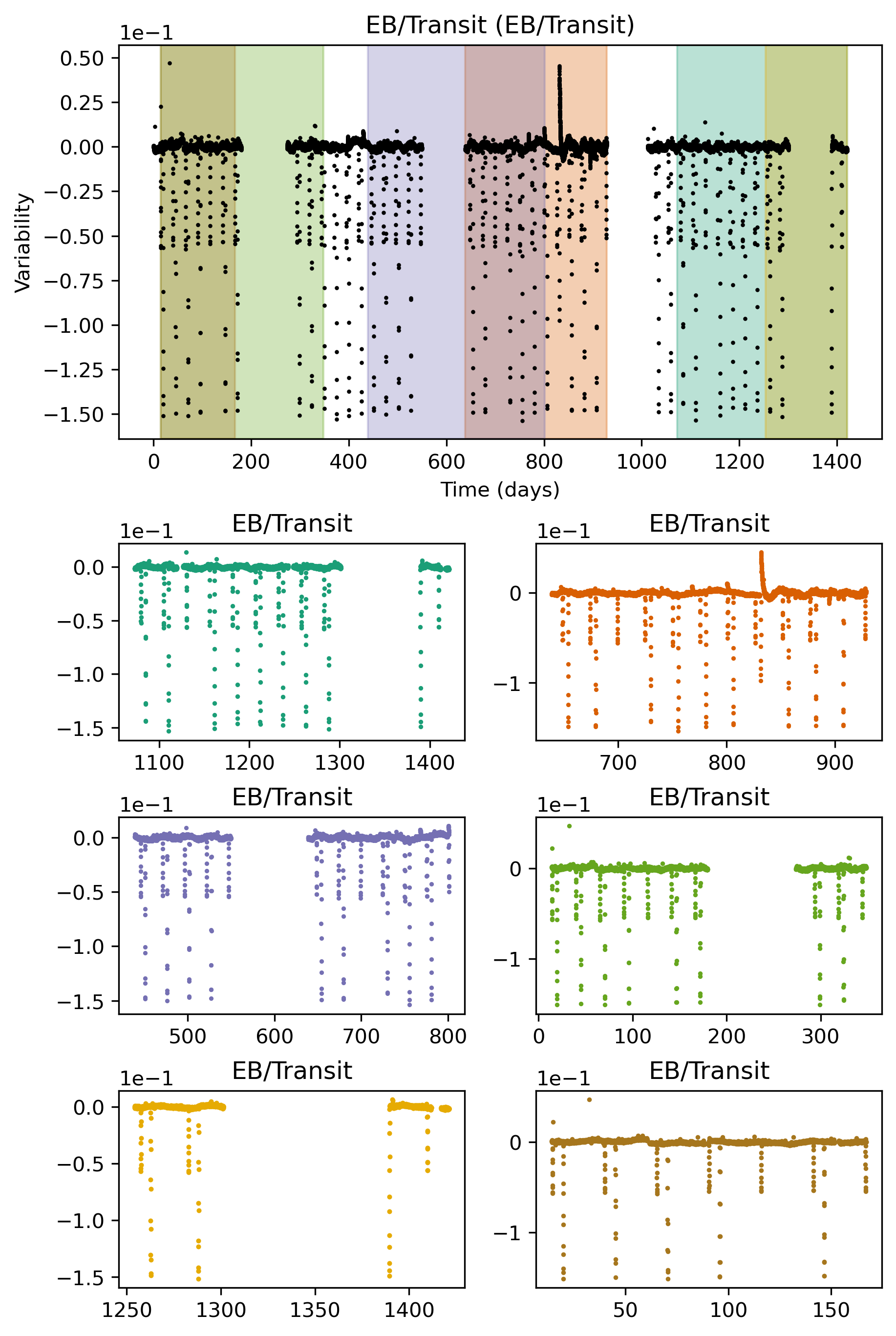}
        \includegraphics[width=0.9\textwidth]{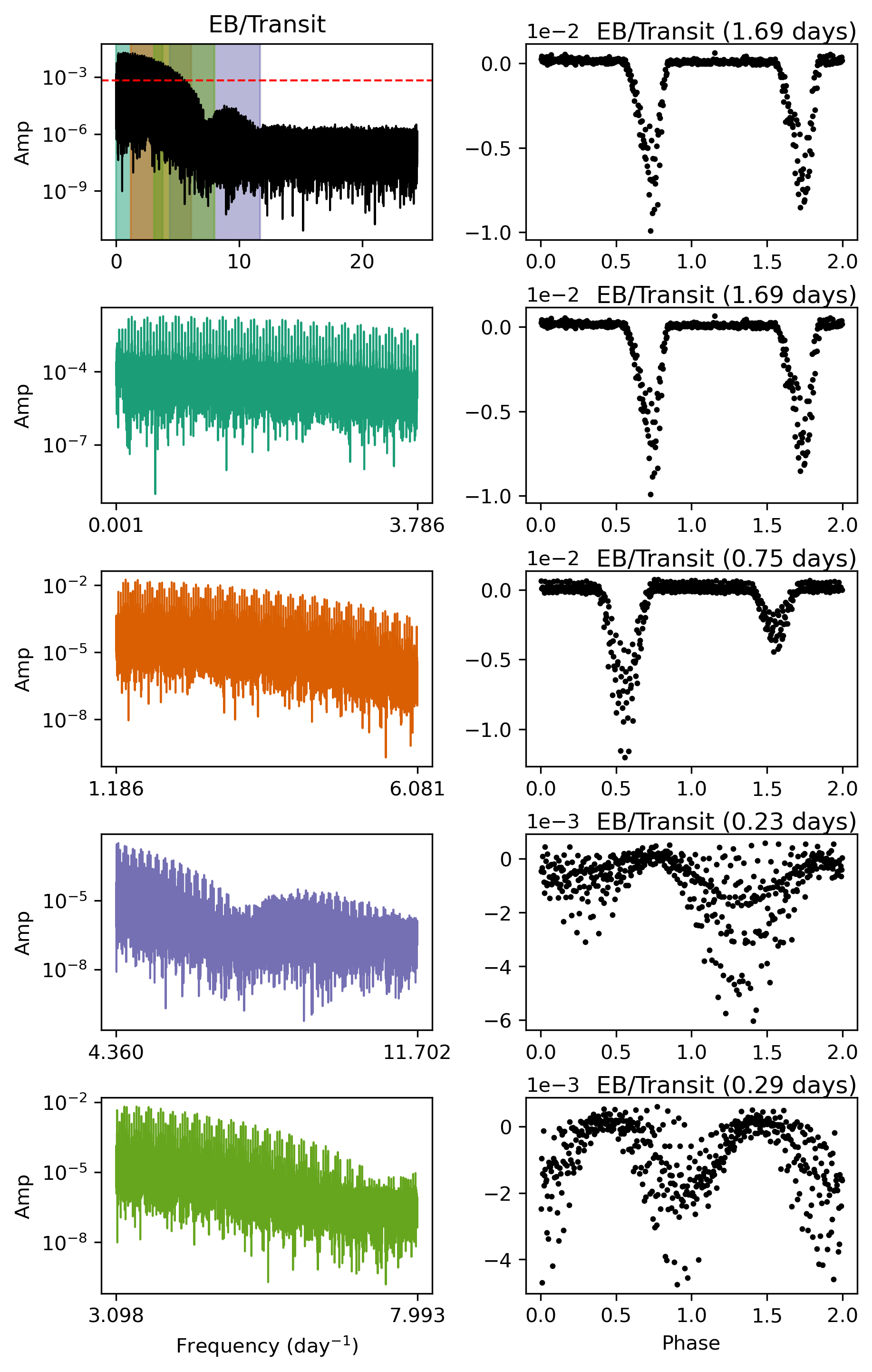}
        \caption{{Intermediate classification results} of KIC 10651945, which is an EB. Similar to Figure \ref{fig: tess-dsct-example}.}\label{fig: Kepler-EB-example}
    \end{minipage}\hfill
    \begin{minipage}{0.49\textwidth}
        \centering
        \includegraphics[width=0.9\textwidth]{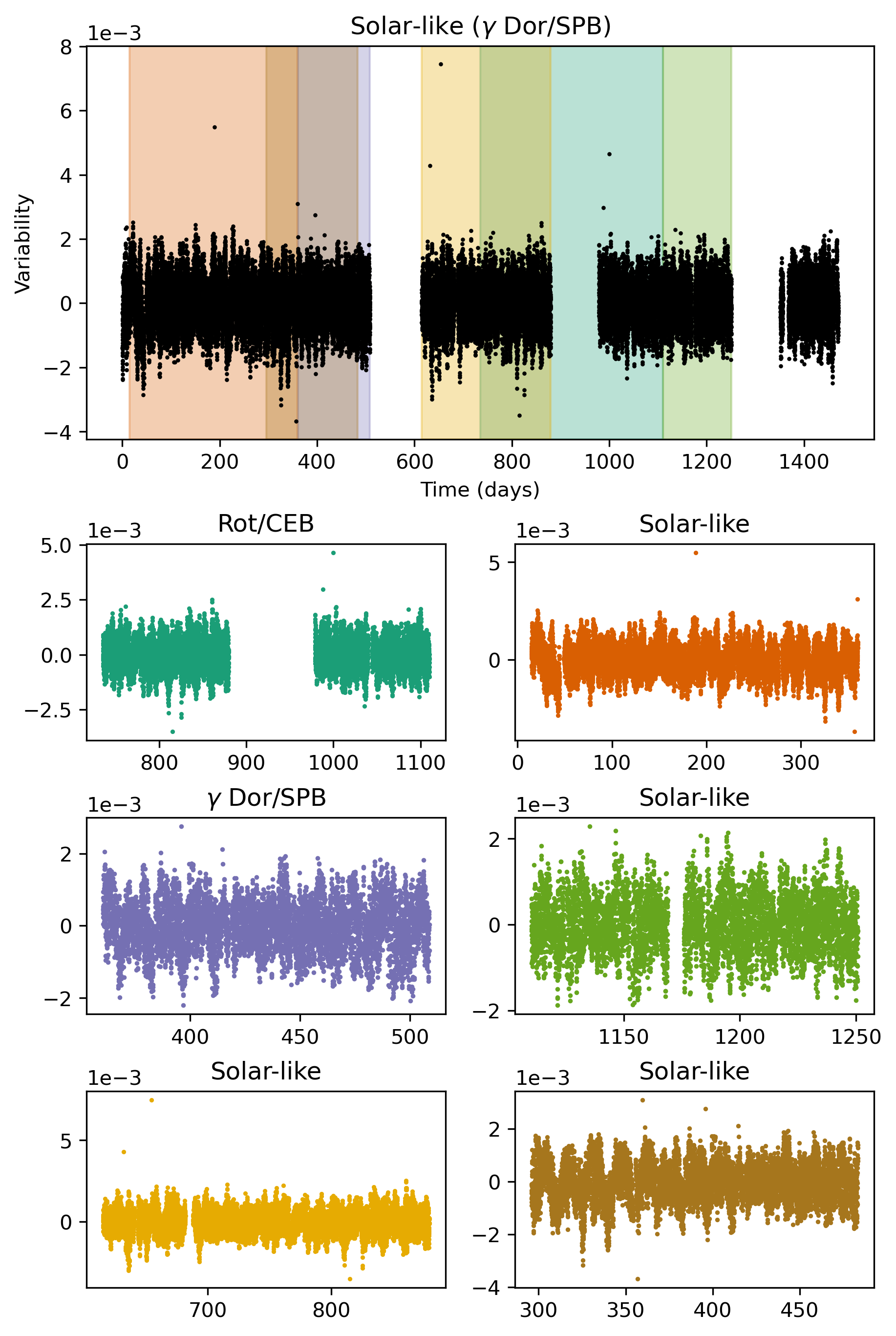}
        \includegraphics[width=0.9\textwidth]{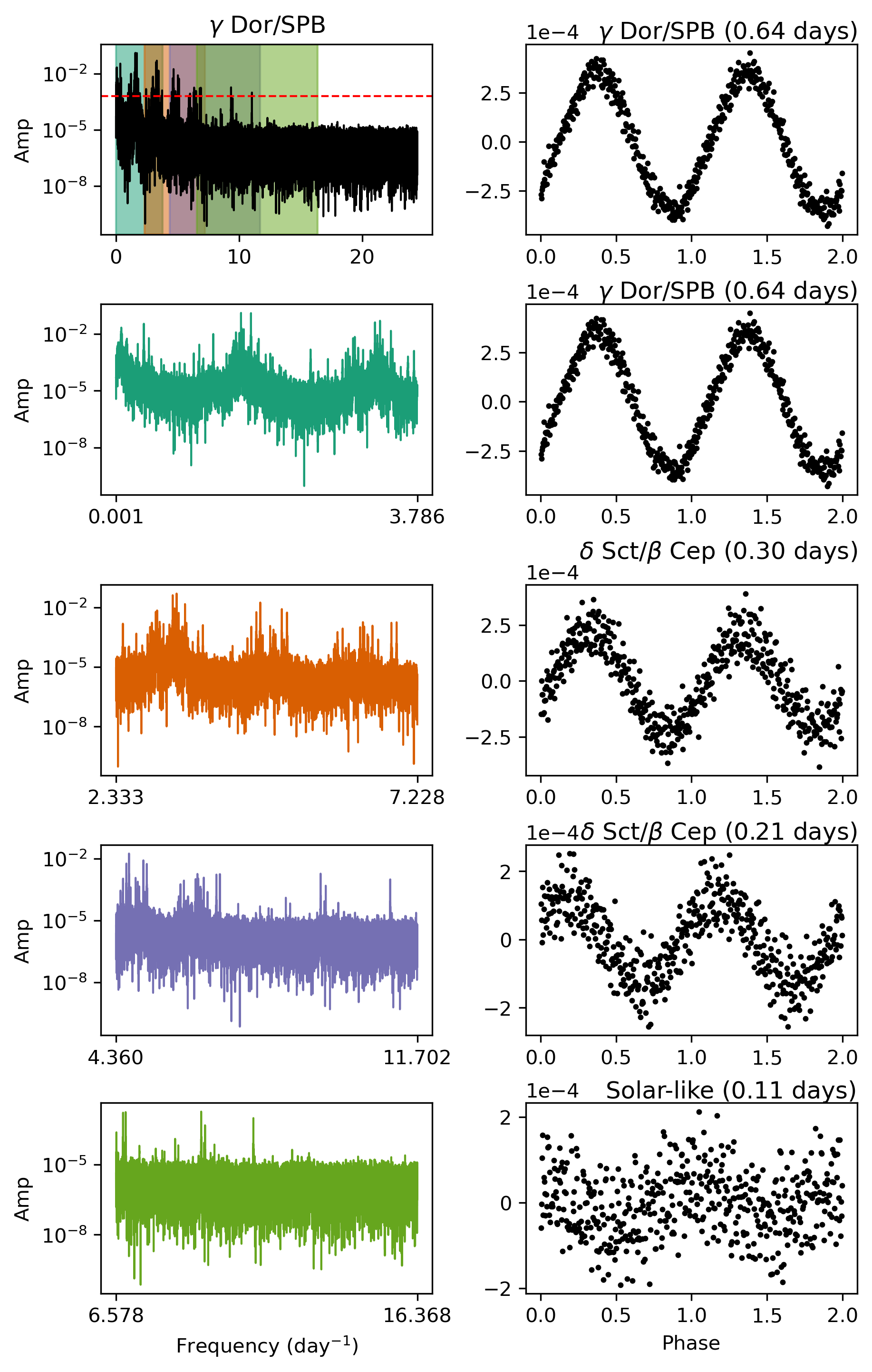}
        \caption{{Intermediate classification results} of KIC 10091792, which is a $\gamma$ Dor. Similar to Figure \ref{fig: tess-dsct-example}.}\label{fig: kepler-gdor-example}
    \end{minipage}
\end{figure}

\bibliography{glcc}
\bibliographystyle{aasjournal}



\end{document}